\documentclass[iop]{emulateapj}
\usepackage{graphicx}
\usepackage{aas_macros}
\usepackage[figuresright]{rotating}
\usepackage{amsmath}
\usepackage{amssymb}
\usepackage{graphics}
\usepackage{wrapfig}
\usepackage{soul}
\usepackage{natbib}
\usepackage{array}
\usepackage{appendix}
\usepackage{lineno}

\usepackage[table]{xcolor}
\usepackage{booktabs,colortbl,tabularx}
\newcolumntype{P}[1]{>{\centering\arraybackslash}p{#1}}
\newcolumntype{M}[1]{>{\centering\arraybackslash}m{#1}}
\usepackage{comment}
\usepackage{lipsum}
\usepackage{threeparttable}
\usepackage{rotating}

\newcommand{\um}{\ensuremath{{\mu}}m\;}

\newcommand{\ha}{{H}{$\alpha$}~$\lambda$6565}
\newcommand{\hb}{{H}{$\beta$}~$\lambda$4863}

\newcommand{\z}{$1<z\lesssim2$\;}

\newcommand{\thot}{$\rm kT_{e,hot}$\;}
\newcommand{\twarm}{$\rm kT_{e,warm}$\;}

\newcommand{\gamwarm}{$\rm \Gamma_{warm}$\;}
\newcommand{\rhot}{$\rm R_{hot}$\;}
\newcommand{\rwarm}{$\rm R_{warm}$\;}
\newcommand{\rout}{$\rm R_{out}$\;}

\newcommand{\maz}[1]{\textcolor{black}{{#1}}}

\def \wilkes {\cite{Wilkes2013}}
\def \elvis {\cite{Elvis1994}}
\def \sieb {\cite{Siebenmorgen2015}}
\def \kd {\cite{Kubota2018}}
\def \dc {\cite{dc2008}}
\def \dcr {\cite{dcr2015}}

\def \m {MAGPHYS\;}
\def \pece {\cite{pece15}}

\newcommand{\reff}[1]{\textcolor{black}{{#1}}}

\def\nar{\ref@jnl{New A Rev.}}          

\shorttitle{Disentangling the AGN and star-formation contributions to the radio-X-ray emission}
\shortauthors{Azadi et al.}
\slugcomment{Accepted to ApJ}

\begin{document}

\title{Disentangling the AGN and star-formation contributions to the radio-X-ray emission of radio-loud quasars at $1<z<2$}

\author{Mojegan Azadi\altaffilmark{1},
Belinda Wilkes\altaffilmark{1}, 
Joanna Kuraszkiewicz\altaffilmark{1}, 
Jonathan McDowell\altaffilmark{1},
Ralf Siebenmorgen\altaffilmark{2},
Matthew Ashby\altaffilmark{1}, 
Mark Birkinshaw\altaffilmark{3},
Diana Worrall\altaffilmark{3},
Natasha Abrams\altaffilmark{8},
Peter Barthel\altaffilmark{4},
Giovanni G.\ Fazio\altaffilmark{1},
Martin Haas\altaffilmark{5},
S\'oley Hyman\altaffilmark{6},
Rafael Mart\'inez-Galarza\altaffilmark{1},
Eileen T. Meyer\altaffilmark{7}}
\altaffiltext{1}{Center for Astrophysics $|$ Harvard \& Smithsonian, 60 Garden Street, Cambridge, MA, 02138, USA}
\altaffiltext{2}{European Southern Observatory, Karl-Schwarzschild-Str. 2, 85748 Garching b. M\"unchen, Germany}
\altaffiltext{3}{H.H. Wills Physics Laboratory, University of Bristol, UK}
\altaffiltext{4}{Kapteyn Institute, University of Groningen, The Netherlands}
\altaffiltext{5}{Astronomisches Institut, Ruhr-University, Bochum, Germany}
\altaffiltext{6}{Steward Observatory, University of Arizona, 933 North Cherry Avenue, Tucson, AZ, 85721, USA}
\altaffiltext{7}{Department of Physics, University of Maryland Baltimore County, Baltimore, MD, 21250, USA}
\altaffiltext{8}{Astronomy Department, University of California at Berkeley, Berkeley, CA 94720, USA}


\begin{abstract}

We constrain the emission mechanisms responsible for the prodigious electromagnetic output generated by active galactic nuclei (AGN) and their host galaxies with a novel state-of-the-art AGN radio-to-X-ray spectral energy distribution model fitting code (ARXSED). 
ARXSED combines multiple components to fit the spectral energy distributions (SEDs) of 
AGN and their host galaxies. Emission components include radio structures such as lobes and jets, 
infrared emission from the AGN torus, visible-to-X-ray emission from the accretion disk, and radio-to-ultraviolet emission from the host galaxy. 
Applying ARXSED to the radio SEDs of 20 3CRR quasars at $1 < z < 2$ verifies the need for more than a simple power law when compact radio 
structures are present. \reff{The non-thermal emission contributes 91\%–-57\% 
of the \maz{observed-frame} 1.25\,mm to 850\,\micron\  flux, and this component must be 
accounted for when using these wavelengths to estimate star-formation properties. \maz{We predict the presence of strong radio-linked X-ray emission in more than half the sample sources.} ARXSED estimates median (and the associated first and third quartile ranges) BH mass of $2.9_{1.7}^{6.0} \times 10^9~\rm M_{\sun}$, logarithm of Eddington ratio of $ -1.0_{-1.2}^{-0.6} $, and spin of $ 0.98_{0.94}^{0.99} $ for our sample. The inferred AGN torus and accretion disk parameters agree with those estimated from spectroscopic analyses of similar samples in the literature. We present the median intrinsic SED of the luminous radio-loud quasars at \z; this SED represents a significant improvement \maz{in the way each component is modeled}.}

\end{abstract}


\section{Introduction} \label{sec:intro}


It is now widely accepted that the center of almost every galaxy
hosts a supermassive black hole (SMBH), which by accretion of gas and dust can turn into an active galactic nucleus 
\citep[AGN; for a review see][]{Heckman2014}. While there has been significant progress in our understanding of galaxy formation and evolution in the past decades \citep[e.g.,][]{Madau2014}, there is as yet no coherent picture of the role AGN play in this evolution \citep[e.g., see][]{Page2012,aird2012primus,Rovilos2012,Rosario2012,Azadi2015,Bernhard2016,Brown2019,Shangguan2020}. As both release a tremendous amount of energy over a wide range of wavelengths, it is critical to disentangle the radiation from the AGN and the host galaxy as a function of wavelength in order to understand better the emission mechanisms operating in them.

\reff{In the standard picture of a powerful active galaxy \citep[e.g.,][]{Urry1995}, a central SMBH is surrounded by a gaseous, optically thick, geometrically thin, accretion disk that generates visible (O), ultra-violet (UV) and soft X-ray emission \citep[e.g.,][]{NT1973,SS1976,Rees1984}. This central engine is, in turn, surrounded by an asymmetric dusty structure known as the torus, with a half-opening angle
of $\sim 60 ^{\circ}$ \citep[e.g.,][]{Willott2000, Wilkes2013}. The torus absorbs some of the ultraviolet (UV) and optical photons and re-radiates them at near-infrared (NIR) to MIR-infrared (MIR) wavelengths \citep[e.g.,][]{Neugebauer1979, Rieke1981}. Radiation from the torus dominates an AGN's emission at 1--40\,\um 
\citep[e.g.,][]{Elvis1994,Netzer2007,Mullaney2011}. A subset of AGN \citep[$\sim15\%$,][]{Kellermann1989,Urry1995}, known as radio galaxies, also emit strongly at radio frequencies \citep[for a recent review see][]{Hardcastle2020} as a result of the interactions of relativistic electrons (or positrons) with the magnetic fields \citep[see][for a detailed review]{GH2013}. Radio galaxies are designated as FRII or FRI based on their morphology: FRII sources are double-lobed and are brightest near the ends of the lobes, while FRIs have the brightest extended radio emission near the center of the structure \citep{FR1974}. As for all AGN, radio galaxies are classified as broad-line (Type 1) or narrow-line (Type 2) depending on the presence/absence of broad UVOIR emission lines, which is a function of viewing angle.} 
\maz{
The present work focuses on a sample of broad-line radio galaxies having FR II morphology and high power that are classified as quasars.}

\reff{Below, we briefly describe the major AGN components according to a selective survey of the recent literature. The elements considered include the AGN accretion disk, the torus, and the radio structures. \maz{Our radio-to-X-ray spectral energy distribution model fitting code (ARXSED) accounts for the emission from all these elements.}}


{\sl The accretion disk:} 
\reff{since the classical models of \cite{NT1973} and \cite{SS1976} there have been many attempts to model the emission from the accretion disk \citep[e.g.][among others]{Done2012,Done2013,Petrucci2013,Kubota2018}}. Generally, the temperature and optical depth of the gas in the disk are inversely related to distance from the SMBH. The temperature gradient results in an energy gradient in the emission, with hard X-rays ($>2$\,keV) originating in the inner region while UV, and visible radiation originate in the outer disk \citep[e.g.,][]{Done2012,Kubota2018}. The hard X-ray radiation is due to the Compton up-scattering of the accretion disk photons in a hot (electron temperatures $\sim$100\,keV) and optically thin corona located above and below the inner disk \citep[e.g.,][]{Haardt1993,Svensson1994,Petrucci2001}. The hard X-ray emission is usually consistent with a simple power-law spectrum. 
At lower energies ($\sim 2$\,keV),
many AGN show a soft X-ray excess above the extension of the harder power-law \citep[e.g.,][]{GD2004, Piconcelli2005, Bianchi2009}. There is no single, widely accepted origin for the soft excess \citep[e.g.,][]{Crummy2006, Czerny2003, Petrucci2018}. \reff{In this study, we adopt the accretion disk model of \cite{Kubota2018}, which assumes that the soft X-ray excess results from Comptonization of thermal photons by a warm ($ kT_{e}\sim$ 0.1--1\,keV), optically thick ($\rm \tau \sim $10--25) layer above the surface of the disk  \citep[also see][]{Czerny2003,Kubota2018,Petrucci2018}. In this model}, the total radiative power of the accretion disk depends on the black hole mass, the mass accretion rate, the spin of the SMBH, and the radial dependence of the optical depth \citep[e.g.,][]{DL2011, Done2012, Kubota2018}.

{\sl The torus:} the distribution of dust in the torus has been the subject of many studies. Early studies proposed a homogeneous structure in which the dust is smoothly distributed in a toroidal disk \citep[e.g.,][]{Pier1992,Efstathiou1995,Fritz2006}. However, these models are not able to describe some of the observed features, for example, the 
9.7\,\um\ silicate absorption feature in Type 1 sources \citep[e.g.,][]{Roche1991}. A significant part of the MIR radiation comes from the polar regions \citep[e.g.,][]{Braatz1993,Honig2013}, which toroidal models could not explain.
A clumpy circumnuclear torus was then put forward as a possible solution \citep[e.g.,][]{Nenkova2008,HK2010,Honig2013}.
Recently \sieb\ proposed a model in which the dust can be distributed in a homogeneous disk, a clumpy medium, or a combination of both \citep[][]{sieb2005,Feltre2012}. \sieb\ model reproduces the MIR spectra of AGN, including the 9.7 \um silicate absorption feature in Type 1 AGN, the radiation from the hot dust close to the sublimation temperature, and the MIR radiation from the ionization cone \citep[e.g.,][]{Braatz1993,Cameron1993,Honig2013}.\maz{ ARXSED incorporates the \citeauthor{Siebenmorgen2015} torus model.}

{\sl The radio structures:} 
the radio-loud AGN launch powerful jets which persist as highly collimated structures until they terminate as bright shocks (``hot spots") at the interface with the circumgalactic or intergalactic medium. 
The observed shape of the radio spectra depends on the \reff{age and acceleration of the electron population, and viewing angle, among other factors.}
In a simplified picture, the radio spectrum of a lobe-dominated AGN can be described with a power-law ($L_{\nu}$ $\propto \nu^{\alpha}$). However, in AGN with compact radio structures such as radio cores and hot spots superposition of various components makes the shape of the spectrum more complex. Therefore, to model the radio emission, multi-component models with different spectral indices are required. The radiation from these structures eventually breaks at high frequencies as a result of synchrotron radiation losses and terminates at a cutoff frequency due to \reff{a drop in the electron population caused by rapid energy losses \citep[e.g.,][]{Blandford1979,Konigl1981}.}

What sparks the radio-loud phase in some AGN is still unknown and may be stochastic by nature. Some studies find that the triggering mechanism of the radio-loud phase is intrinsic to the AGN rather than the host galaxies or the environments \citep[e.g.,][]{Kellermann2016,Coziol2017}, while others find cold star-forming gas in galaxies \citep[e.g.,][]{Janssen2012,BH2012} or the brightness of the cluster and density of the environment they live in 
\citep[e.g.,][]{Burns1990,Best2007} increase the likelihood of hosting a radio-loud AGN.  The prevalence of radio-loud AGN and the power of their radio emission (i.e., $ L_{\rm 1.4 GHz}$) correlates strongly with the intrinsic properties of the SMBH 
\citep[e.g., SMBH mass,][]{Best2005,Coziol2017}. 
\reff{Some studies suggest that the BH mass is a critical parameter in dividing the radio-loud and radio-quiet populations \citep[e.g.,][]{Chiaberge2011}. There is as yet no consensus on whether the Eddington ratio (which is the ratio of the bolometric luminosity to the Eddington luminosity)
\begin{align}
\label{eq:edd}
  &\lambda_{Edd} = L_{bol}/L_{Edd}
  \\
  &L_{Edd} \propto M_{BH} \nonumber
\end{align}
and/or spin has a role in triggering the radio loud phase \citep[e.g.,][]{Sikora2007,Chiaberge2011,Coziol2017}.}


The multiple components described above emit radiation covering 10 decades of frequency from X-ray to radio. To constrain these components, we, therefore, need observations across a broad range of wavelengths. However, obtaining a multi-wavelength dataset is very challenging, and in the case of quasars, variability adds more complications. Over the past three decades, numerous studies have focused on spectral energy distribution (SED) analysis of AGN populations \citep[e.g.,][]{Edelson1986,Elvis1994,Richards2006,shang2011,Elvis2012,Hao2014}. 
A number of these studies attempt to understand whether quasars' behavior can be described with an average SED. \elvis\ presented the first high-quality broad (X-ray to radio) atlas of quasar SEDs at $z \lesssim 1$ using then-current telescopes such as \textit{Einstein}, the \textit{International Ultraviolet Explorer (IUE)}, and \textit{ Infrared Astronomical Satellite (IRAS)}. \cite{Elvis1994} presented median SEDs for the radio-loud and radio-quiet quasars which have been extensively used and overall work remarkably well in the 0.1--1 \um wavelength range for AGN of various luminosities and Eddington ratios \citep{Elvis2010}. However, there is a large dispersion around their median SED which can reflect on the inferred properties of the quasars. Additionally, their sample is not representative of the overall quasar population and is biased towards X-ray bright and blue quasars \citep[e.g.,][]{Jester2005,shang2011}.

The average SED of quasars has been investigated by others in a similarly broad range of the spectrum, radio-to-X-ray, \citep[e.g.,][]{ Richards2006,shang2011,Elvis2012} or a more limited range \citep[e.g.,][among others]{JK2003,Polletta2007,Mullaney2011,Lani2017} using more recent spectroscopic and photometric data. \cite{Richards2006} found a wide range of SED shapes for quasars and concluded that assuming an average SED can potentially lead to 50\% errors in the bolometric luminosity estimate. \cite{Hao2014} found that in X-ray selected, radio-quiet quasars the average SED has no significant dependence on redshift, bolometric luminosity, SMBH mass, and Eddington ratio, while others \citep[e.g.,][]{Geach2011,Kirkpatrick2017} found that local quasar templates may not be applicable to AGN at higher redshifts
. One of the many challenges in AGN SED analysis is the contamination of the broadband photometry by the host galaxy. Most of the studies noted here use scaling relations such as $\rm M_{BH}-\sigma$ \citep[e.g.,][]{marconi2003relation} or $\rm  L_{host}-L_{AGN}$ \citep{Vanden2006} or color-color diagnostics \citep[e.g.,][]{Grewing1968,Sandage1971} to estimate the host galaxy contamination. However, each of these scaling relations has large uncertainties that significantly affect the SED.


With the wealth of information stored in the photometric data obtained with modern telescopes with higher sensitivity and resolution, many studies have moved towards fitting complex models to the broadband photometry that can describe the radiation from the AGN and the host galaxy simultaneously and provide estimates of the physical properties of each (i.e., SMBH mass, stellar mass). The galaxy SED models (e.g., MAGPHYS:  \citealt{dc2008}; FAST: \citealt{kriek2009}; CIGALE: \citealt{Boquien2019}) adopt libraries of galaxy templates with different stellar populations. These libraries are often built on assumptions on star formation history (SFH), initial mass function (IMF), dust attenuation law, etc. which each has their own uncertainties. In the case of active galaxies, additional templates (i.e., torus, accretion disk) are required to describe the radiation from the central engine. For instance, \cite{Berta2013} presented the SED3fit model, which combines the galaxy templates from MAGPHYS \citep{dc2008} with the torus templates of \cite{Fritz2006}
to describe the radiation from the AGN and its host within 8-500 $\rm \mu m$. \cite{Leja2018} use the galaxy SED model of Prospector-$\alpha$ \citep{Leja2017} and the clumpy torus model of \cite{Nenkova2008} to fit the SED of nearby AGN at 1-100 $\mu m$. While overall these models are successful in modeling AGN and the host galaxy's emission, portraying a complete picture of AGN requires a dataset that covers the radiation from the accretion disk, which is the primary source of AGN power, and the radio component (in the case of radio-loud AGN). 

\reff{The present work describes SED modeling of a subset of AGN from the Revised-Third Cambridge Catalogue of Radio Sources \citep[3CR;][]{Spinrad1985}.  Specifically, we construct the radio to X-ray SEDs of 3CRR\footnote{The Revised-3CR (3CRR) is the most complete version of the 3CR sample that only includes extragalactic sources,
\reff{with 178-MHz flux density limit} $>10$Jy, declination $>10^{\circ}$, and Galactic latitude $>10^{\circ}$ or $<-10^{\circ}$.} \citep{Laing1983} quasars at \z, which together with narrow-line radio galaxies in the same redshift band define an orientation-independent sample of luminous AGN.  We focus on this epoch -- known as Cosmic High Noon -- because
of its well-known significance for the growth of SMBHs and the assembly of the galaxy bulges in which they reside. SED fitting of the full IR continuum of 3CR sources at a range of redshifts indicates that orientation determines the MIR--NIR continuum shapes \citep[e.g.,][]{Haas2008,Drouart2014,pece15,pece2016}.  Here we expand on earlier SED fitting of the IR continuum to include the radio-to-X-ray emission.
This paper introduces the detailed, state-of-the-art AGN Radio-X-ray SED modeling code (ARXSED) and applies it to
the 3CRR quasars.  We present the estimated physical properties of the AGN components derived from the modeling (e.g., SMBH mass, Eddington ratio, and spin). SED fitting of the narrow-line radio galaxies and properties of the host galaxies will be presented in future papers.}

 This paper is organized as follows: Section 
\ref{sec:data} describes the sample selection and the multi-wavelength data used for the SED analysis. Section \ref{sec:sed} presents the details of our SED modeling and the components used to describe the emission from the AGN and the host galaxy at different wavelengths.
Section \ref{sec:fitting_analysis} describes the fitting results for individual sources, and results of the SED analysis are discussed in Section \ref{sec:discussion}. We present a summary of our findings in Section \ref{sec:summary}. Throughout the paper we adopt a flat cosmology with $\Omega_{\Lambda}$ = 0.7 and $H_0$= 72 km s$^{-1}$ Mpc$^{-1}$.

\section{Sample and Data Compilation}\label{sec:data}

\reff{The 3CRR catalog \citep{Laing1983} includes 180 FRII radio galaxies and quasars 
up to a redshift of 2.5 and is 100\% complete to a 178\,MHz flux density of 10\,Jy.} \maz{Of these, 38 sources are at \z \citep{Wilkes2013} including 21 that have broad emission lines (and classified as quasars), and 17 showing only narrow emission lines.}
\reff{In this study, we limit our sample specifically to the quasars to ensure that the accretion disk dominates the visible-UV radiation. The sample is presented in Table \ref{tab:sample}. We have excluded the red quasar (3C~68.1) from our analysis as its SED is similar to those of the narrow-line radio galaxies. 3C~68.1 will be discussed in a future paper.}

AGN have obscuration-dependent emission that results in strong selection effects at most wavelengths \citep[e.g.,][]{Azadi2017}. \reff{However, selection based on optically-thin, low-frequency (178 MHz) radio emission uniquely finds AGN without the orientation bias. The complete nature of the 3CRR catalog, combined with the sources' high brightness and luminosities, freedom from orientation bias, and the availability of comprehensive multi-wavelength data, makes the 3CRR sources an optimal sample to characterize the luminous radio-loud AGN population.}

The radio properties of our sample, from the radio morphologies and the lengths of the radio structures (e.g., radio extents measured from lobe to lobe) to the fraction of radiation from the radio core relative to the lobes at 5 GHz (a jet orientation indicator), among many other properties, have been studied in great detail \citep[][]{Akujor1991,Bridle1994,Akujor1995,Lonsdale1984,ODea1998,Ludke1998}. 

Although the 3CRR sample has been extensively studied, surprisingly, many sources miss critical data needed for estimating the basic parameters of their central engines and their host galaxies. For the SED analysis, we have compiled the X-ray to radio SED by combining archival data from \textit{Chandra}, \textit{XMM-Newton}, {SDSS}, {UKIRT}, {2MASS}, \textit{Spitzer}, \textit{WISE}, \textit{Herschel} and multi-frequency radio observations. \reff{The photometry is aperture corrected.} The X-ray data used in this analysis are from \wilkes\, and the optical and UV data are gathered from the references available on NASA/IPAC Extragalactic Database (NED). The NIR to FIR photometry used here is drawn from \pece. Table~\ref{tab:submmdata} indicates the submm fluxes recently obtained from the SMA or ALMA for our analysis. Tables~A1--A3 in the Appendix present the radio observations (with their references) used in our analysis. 

We assembled the final SEDs from these datasets by selecting the highest-quality photometry. \reff{Photometric measurements that showed significant deviations from the majority in that wavelength range were removed.} We excluded photometry with S/N$<3$ at any wavelengths except at FIR and submm wavelengths, where the upper limits are the only constraints available.  
Finally, we note that the visible-UV photometry is not contaminated by emission lines. We fitted a power-law continuum in IRAF \citep{Doug1986} to regions of the \reff{SDSS or HST} spectrum that were uncontaminated by emission lines using the continuum windows of \cite{JK2002}. The power-law was
then binned into 10--15 continuum bins and included in our SEDs. \maz{The spectra and photometry are both dereddened for Galactic absorption (see Section \ref{sec:dered}) for more details.}

\begin{table}[]
    \centering
        \caption{The 3CRR quasars modeled in this work}
    \begin{tabular}{ c c c c}
    \hline \hline \vspace{1pt}
         Name  & RA,DEC  & $z$ \\
         &(J2000.0) &\\ 
         \hline 
\rule{0pt}{2.5ex}3C 009 & 00:20:25.2,+15:40:55 &  2.009  \\
3C 014 & 00:36:06.5,+18:37:59 &  1.469  \\
3C 043 & 01:29:59.8,+23:38:20  & 1.459 \\
3C 181 & 07:28:10.3,+14:37:36 &  1.382 \\
3C 186  & 07:44:17.4,+37:53:17  & 1.067  \\
3C 190 & 08:01:33.5,+14:14:42  & 1.195  \\
3C 191  & 08:04:47.9,+10:15:23 &  1.956 \\
3C 204 & 08:37:44.9,+65:13:35 &  1.112  \\
3C 205  & 08:39:06.4,+57:54:17 &  1.534  \\
3C 208 & 08:53:08.8,+13:52:55 &  1.110  \\
3C 212  & 08:58:41.5,+14:09:44 &  1.048  \\
3C 245  & 10:42:44.6,+12:03:31 &  1.029  \\
\;\;\;3C 268.4 & 12:09:13.6,+43:39:21 &  1.398 \\
\;\;\;3C 270.1  & 12:20:33.9,+33:43:12 &  1.532 \\
3C 287  & 13:30:37.7,+25:09:11  & 1.055  \\
3C 318 & 15:20:5.40,+20:16:06  & 1.574  \\
3C 325 & 15:49:58.4,+62:41:22 &  1.135  \\
\;\;\;4C 16.49 & 17:34:42.6,+16:00:31 &  1.880 \\
3C 432  & 21:22:46.2,+17:04:38 &  1.785 \\
\;\;\;3C 454.0& 22:51:34.7,+18:48:40  & 1.757 \\
\hline
    \end{tabular}
    \label{tab:sample}
    \end{table}

or whatever.

\begin{table}[] 
    \centering
       \caption{Recent submm data obtained with the SMA or ALMA}
    \begin{tabular}{ l l l l}
    \hline \hline  \vspace{1pt}
         Name  & $\nu$ (GHz) & Flux density (mJy) & Telescope  \\
         \hline 
\rule{0pt}{2.5ex}3C $212^{1}$ & 98 & $106\pm 5$ &ALMA  \\
3C $212^{1}$ & 233 & $41\pm 2$ &ALMA \\
3C $245^{2}$ & 224 & $84\pm1$ &SMA \\
3C $270.1^{3}$ & 235 & $13.5\pm0.3$ &SMA  \\
3C $318^{4}$ & 315  & $2.3\pm0.6$ &ALMA \\
3C $454.0^{4}$ & 315 & $10.6\pm1.4$ &ALMA \\
\hline
    \end{tabular}
    \label{tab:submmdata}
        \begin{tablenotes}
 
\item[1] 1) PI: Meyer; Proposal ID: 2019.1.01709.S  
\item[2] 2) PI: Ashby; Proposal ID: 2019B-S034
\item[3] 3) PI: Ashby; Proposal ID: 2019A-S031
\item[4] 4) PI: Podigachoski; Proposal ID: 2015.1.00754.S

\end{tablenotes}
   \end{table}

\begin{figure*}[t]
\center
    \includegraphics[width=0.57\textwidth,angle=90]{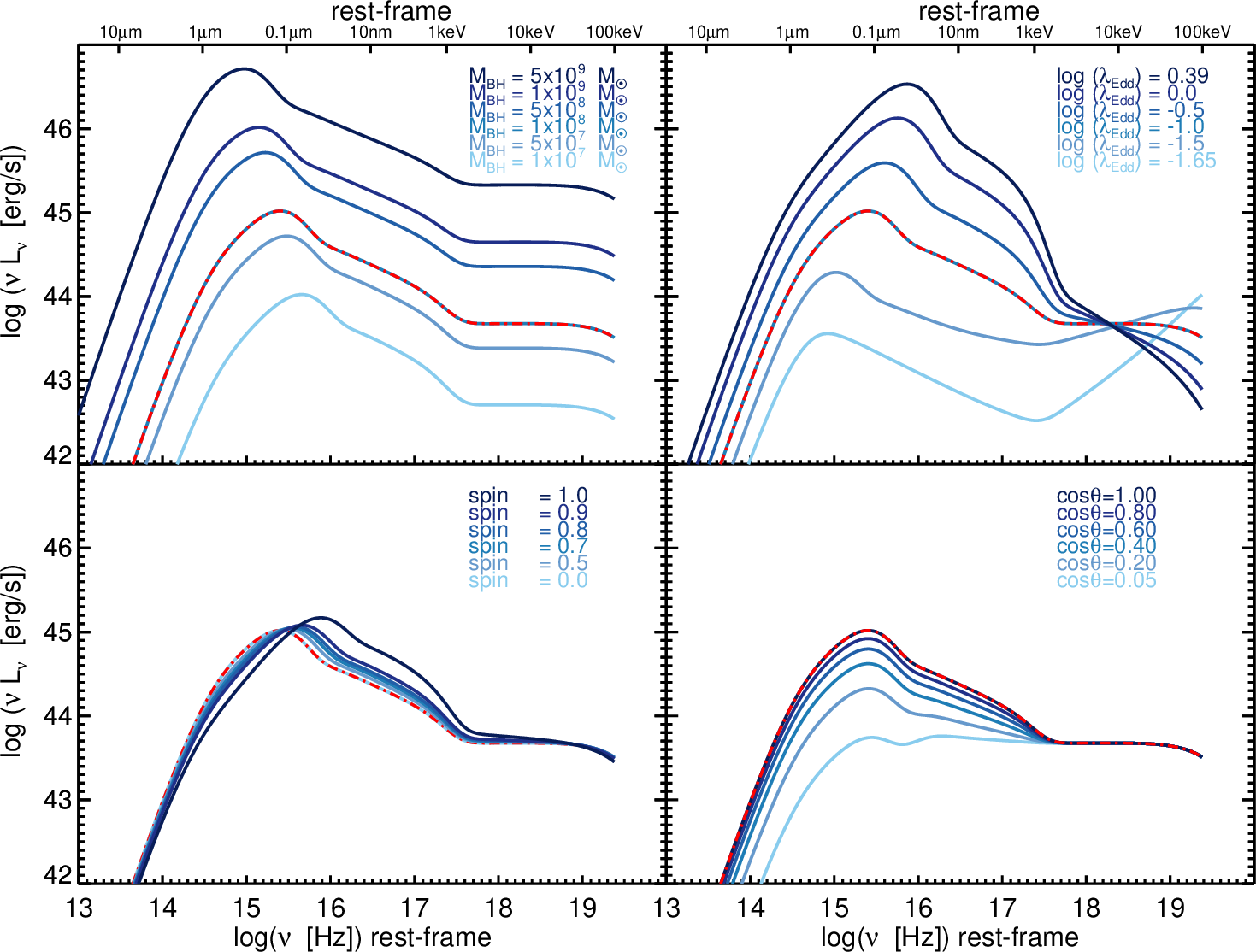}
    \caption{\maz{QSOSED model parameters \citep{Kubota2018} as a function of rest-frame frequency and wavelength.  Clockwise from top left, the parameters varied are BH mass, Eddington ratio, spin, and the inclination angle.  In each panel, the red dotted dashed line indicates the reference model in which $\rm M_{BH} =10^{8} M_{\sun}$, $\log (\lambda_{Edd}) = -1.0$, spin = 0, and $\rm cos \rm \theta = 1$.  Additional curves illustrate the impacts of varying the indicated parameter while holding all other parameters constant.}}
    \label{fig:QSOSED}
\end{figure*}

\begin{table*}[] 
    \centering
       \caption{\reff{The parameters used in constructing the accretion disk templates with the QSOSED model.}}
    \begin{tabular}{ l l l l l}
    \hline \hline \vspace{1pt}
         Parameter  & Initial estimate & Range of variation & Steps & Acceptable range \\
         \hline  
\rule{0pt}{2.5ex}$\rm log(M_{BH})$& $\rm M_{0}$ ({\ion{C}{4}} or {\ion{Mg}{2}})  & [$\rm log(M_{0})$-1,$\rm log(M_{0})$+1] &0.2 dex  &[7,10]\\
$\rm log(\lambda_{Edd})$& -- & $[-1.65,0.39]$ &0.2 dex &[-1.65,0.39] \\
spin (1) & -- & $[0.75^{a},0.998]$ &0.02 &[0,0.998] \\
inclination angle ($\rm \theta_{D}$) & $\rm \theta_{T}(torus)$& [$\rm \theta_{T}$-12,$\rm \theta_{T}$+12] & $\rm \theta_{T}$ dependent &[0,90]  \\

\hline
    \end{tabular}
        \begin{tablenotes}
\item[1] 1) Templates with lower spin values (0-0.75) was examined for a subsample of our sources. However, none of those templates resulted in a better accretion disk fit.
        \end{tablenotes}
    \label{tab:qsosed}
   \end{table*}
 
\section{SED Components and the Fitting Routine} \label{sec:sed}

One of the most important advantages of this study relative to previous work is that it accounts for emission over ten decades in frequency, from radio to X-ray. Because both AGN and host galaxies contribute throughout the electromagnetic spectrum, it is critical to disentangle the contributions from each as a function of wavelength in order to quantify the contributions of the various physical mechanisms to the overall energy budget. In this section, we first describe the components used in our SED model (ARXSED)
to account for the AGN emission, and then describe those used to model the emission from the host galaxy. Additionally, we describe the methods used to correct for the obscuration from the torus, the host galaxy, and the Milky Way absorption along the line of sight.


\subsection{The AGN Components in ARXSED} \label{sec:agn}

Here we treat the models accounting for the AGN emission as arising from three components: an accretion disk, an obscuring torus, and radio lobes.


\subsubsection{The Accretion Disk Component} \label{sec:AD}

For the accretion disk (AD), we use the QSOSED model developed by \kd. The primary variable parameters in this model are presented in Table \ref{tab:qsosed}. 
\reff{Adopting a geometrically
thin disk, } \kd\ assume that the emission from the accretion disk originates in three distinct regions:
an inner region extending from the innermost stable circular orbit ($\rm R_{ISCO}$) to $\rm R_{hot}$, an intermediate region extending from $\rm R_{hot}$ to $\rm R_{warm}$, and an outer region from $\rm R_{warm}$ to $\rm R_{out}$.
AD radii \rwarm and \rhot, respectively define the boundaries of warm and hot Comptonization regions, and $\rm R_{out}$ is the outer edge of the AD 
\citep [for more details see Appendix A of ][]{Kubota2018}.
In this framework, the inner region has a temperature of $\sim$ 100 keV
and includes the hot corona with no underlying disk (i.e., a truncated disk). The plasma in the inner region emits the X-ray power-law component. In the intermediate region, warm Comptonization occurs, and the soft X-ray excess is produced. The electron temperature in the intermediate region is $\sim$0.2 keV, and the optical depth is $\sim10-25$ 
\citep[e.g.,][]{Czerny2003, GD2004, Petrucci2013, Middei2018}. The nature of
the warm Comptonization is not completely understood   \citep[e.g.,][]{Done2012}.  It may result from a failed, UV-driven wind arising in the outer disk region that falls back down into the disk \citep{Laor2014}. The outer disk region, which has a temperature of a several thousand kelvins, is the standard optically thick
AD dominated by blackbody emission. 

The QSOSED model fixes the accretion disk parameters to typical AGN values of \thot = 100\,keV, \twarm = 0.2\,keV, \gamwarm = 2.5, \rwarm= 2\rhot, \rout = $\rm R_{self\,gravity}$ which define respectively the electron temperatures for the hot and warm Comptonization components, the spectral index of the warm Comptonization component, and the radii of the regions.  
The spectral index of the hot  Comptonization component, $\rm \Gamma_{hot}$,  is calculated via equation 6 of \cite{Kubota2018}. 
\reff{The QSOSED model also includes reprocessed radiation, which is the fraction of the hot Comptonization component's emission that 
heats both the warm Comptonized material and the cool outer disk.} 

The primary variables in the QSOSED model are the supermassive black hole mass $\rm M_{BH}$, the mass accretion rate (which is traced with the Eddington ratio defined in equation \ref{eq:edd}), the inclination angle, and the dimensionless spin parameter $\rm a\equiv J\,c/GM_{BH}^{2}$, where $\rm J$ is the angular momentum of the BH.
Figure \ref{fig:QSOSED} illustrates the effect that variation of each parameter has on the shape of the visible, UV, and X-ray SEDs. Overall, the effects of varying these parameters on the geometry of the accretion disk and on its SED are intertwined. 
For example, increasing BH mass while holding mass accretion rate and other parameters constant results in a more luminous but cooler accretion disk.  Increasing the Eddington ratio results in a more luminous and hotter accretion disk. Increasing the spin moves the $\rm R_{ISCO}$
closer to the SMBH and the peak towards higher energies and increases the radiative efficiency. 
\maz{As spin increases toward maximally rotating its impact on the SED shape becomes more noticeable, but lower spins have no significant impact on the peak and shape of the accretion disk SED.} In all cases, a face-on observer sees more of the UV bump than an edge-on observer.

\begin{figure*}[th!] 
         \begin{center}
    \includegraphics[width=0.75\textwidth ,angle =90]{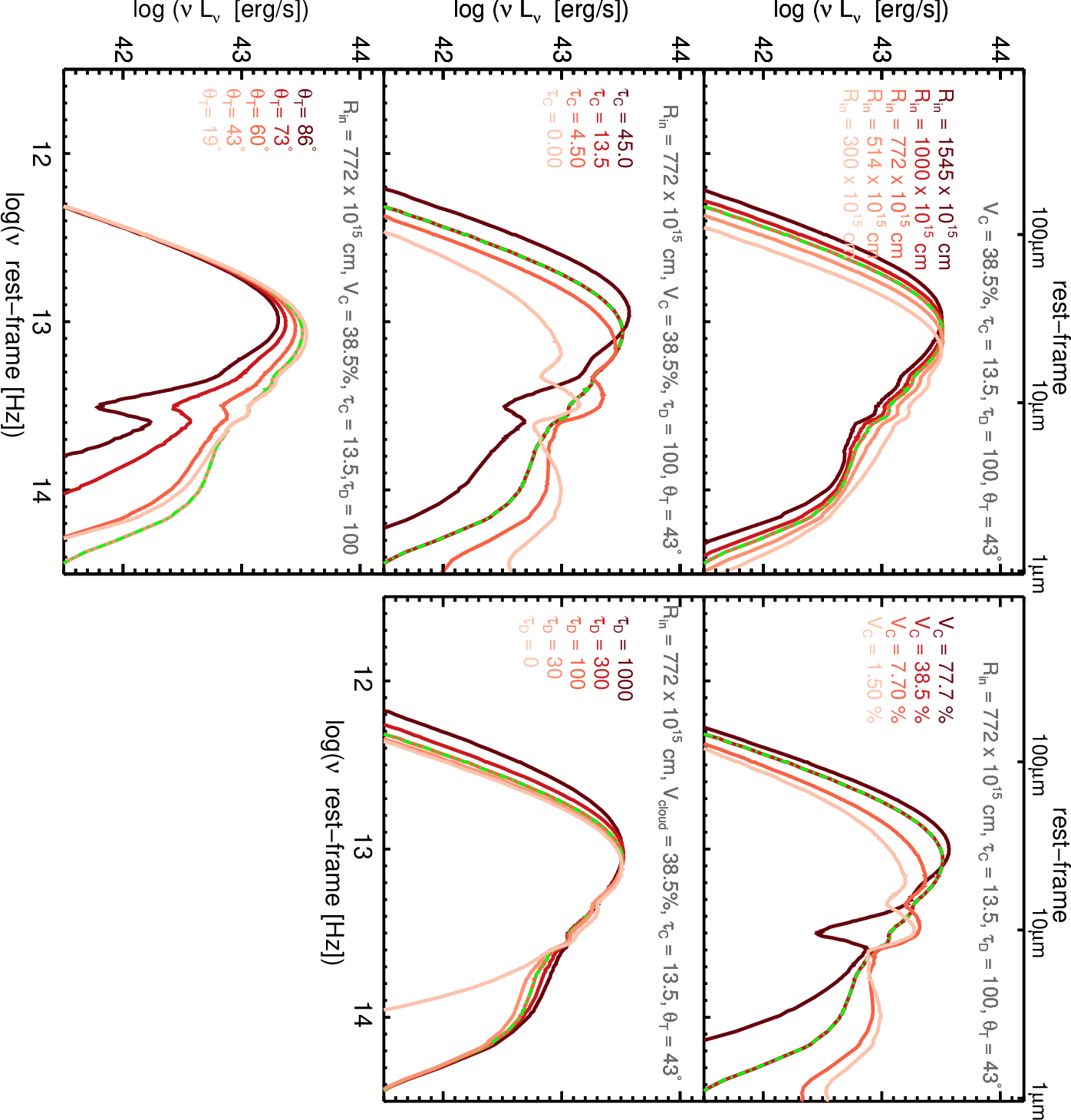} 
         \end{center}
        \caption{\reff{An illustration of the impacts of varying one parameter of the torus SED model 
while holding all other parameters constant. The parameters include the inner radius of the torus ($\rm R_{in}$), the volume filling factor of the clumps and their optical depth ($\rm V_{C}$ and $\rm \tau_{C}$), the optical depth of the homogeneous disk ($\rm \tau_{D}$), and the inclination angle. \maz{The green dotted dashed line shows the reference model in which  $\rm R_{in} =772\times10^{15}$ cm, $\rm V_{C}  = 38.5\%$, $\rm \tau_{C}  = 13.5\%$, and $\rm \tau_{D} = 100$, and $\theta_{T} = 43^{\circ}$.}}}
         \label{fig:torus}
\end{figure*}
   \begin{table*}[ht] 
 \centering
\caption{The free parameters and their possible values in the \sieb\ torus library}
    \begin{tabular}{l l}  \hline \hline
       Parameter   & Acceptable values \\
       \hline 
        \rule{0pt}{2.5ex}Inner radius$^{1}$($\rm R_{in})$ &  300, 514, 772, 1000, 1545 ($\times10^{15} $cm) \\
        Cloud volume filling factor ($\rm V_{C}$)&1.5, 7.7, 38.5, 77.7 (\%) \\
      Cloud optical depth ($\rm \tau_{C}$)  & 0, 4.5, 13.5, 45 \\
      Disk optical depth ($\rm \tau_{D}$)  &  0, 30, 100, 300, 1000\\
      Torus inclination angle ($\rm \theta_{T}$) & 19, 33, 43, 52, 60, 67, 73, 80, 86 ($^{\circ}$) \\
      \hline
    \end{tabular}
    \label{tab:sieb_par}

    \begin{enumerate}
    \item \reff{The inner radius is the dust sublimation radius and scales with the total AGN luminosity as $\rm R_{in}\propto \sqrt L_{AGN}$. The values listed here are for an AGN of a luminosity of $\rm 10^{11}L_{\sun}$.}
    \end{enumerate}
\end{table*}

To fit the visible-to-X-ray SED of each quasar, we construct a set of QSOSED templates in XSPEC\footnote{XSPEC is an interactive program used to fit spectral models to data from visible to gamma-rays. For more information, see https://heasarc.gsfc.nasa.gov/xanadu/xspec/} covering a range for each parameter and using any available information to set the initial parameter values (e.g., BH mass). To construct the templates for each quasar, we varied each parameter with a fixed step within their acceptable ranges in QSOSED. Table \ref{tab:qsosed} summarizes the initial estimates, range of  variation, steps used in our procedure, and the acceptable range for each parameter in the model.

\reff{Out of 20 sources in our sample, six (3C~9/191/205/270.1/432/454.0) have a broad  {\ion{C}{4}~$\lambda$1548} line and six (3C~14/181/186/204/245/268.4) have a broad 
{\ion{Mg}{2}~$\lambda$2800} line
\citep[SDSS archive, priv.\ comm.\ or ][]{Barthel1990} which we
used to estimate the BH masses. For the remaining quasars, we used the median BH mass of these 12 sources 
as the initial mass estimate. BH mass measurements based on the high ionization {\ion{C}{4}} line have larger errors than those estimated from the {\ion{Mg}{2}} or Balmer lines \citep[e.g.,][]{Sulentic2002,Baskin2005,Shen2013}\reff{. However, we used this mass only as the initial estimate and allowed $\rm M_{BH}$ to vary within $\pm$ 1 dex of this value (in steps of 0.2 dex.)} Because there is no prior information on $\rm \log \dot{m}$ we allowed it to vary over the entire acceptable range (in steps of 0.2 dex). Radio-loud quasars are expected to have high spin values \citep[see][and references therein]{Reynolds2019} so we used a limited range from 0.75\footnote{\reff{We tested lower spin values (0-0.75) for a subsample of our sources. However, none of those templates resulted in a better accretion disk fit.}} to 0.998 (in steps of 0.02). For accretion disk's inclination angle, we used the best-fit value from the torus model (Section \ref{sec:torus}) and then varied the inclination angle within $\pm 12^{\circ}$. However, rather than using a fixed step size, we determined seven values within $\pm 12^{\circ}$ of the initial estimate empirically \footnote{\reff{We note that the viewing angle from the torus was used as an initial estimate to build the accretion disk templates. However, studies support the connection between the torus and the accretion disk inclination angles \citep[among other parameters, including the Eddington ratio][]{Calderone2012,Campitiello2021}.}}. Overall, we constructed $\sim$ 11000 templates to fit the visible-X-ray SED of each quasar.}

%



 \subsubsection{The Torus Component} \label{sec:torus}
 
We adopted the two-phase AGN torus model developed by \sieb. 
The model assumes the dust around the AGN can be described as a clumpy medium, a homogeneous disk, or a combination of both. \sieb\ defined the inner radius ($\rm R_{in}$) of the dusty structure by the sublimation temperature of the dust grains. The outer radius of the structure is
 $\rm R_{out}= R_{in} \times 170$, which is chosen to be large enough to have a negligible impact on the observed features in the FIR SED.
The density of the dust decreases with radius.
The dust grains are fluffy mixtures of silicate and amorphous carbon grains \citep{Krugel1994} instead of the interstellar medium (ISM) dust grains \citep[e.g.,][]{Siebenmorgen2014}. These fluffy grains are more efficient absorbers and have higher submm emissivities than the diffuse ISM grains \citep[for more details, see Section 2.6 of][]{Siebenmorgen2015}.

Table~\ref{tab:sieb_par} lists the primary  variable  parameters and their possible values in the model. These parameters are the inner radius of the dusty structure ($\rm R_{in}$), the optical depth of the homogeneous disk mid-plane ($\rm \tau_{D}$), the optical depth of the clumps ($\rm \tau_{C}$), the volume filling factor ($\rm V_{C}$), and the torus inclination angle ($\rm \theta_{T}$). Varying these parameters within their acceptable range makes a library with 3600 templates. These templates are built for an AGN of luminosity $\rm 10^{11}L_{\odot}$. \reff{Since the SEDs are scale-invariant, the inner radius scales (see Table 6) with the square root of the actual luminosity of the primary source, i.e., the accretion disk, 
\citep[e.g.,][]{Suganuma2006,Kishimoto2011}.}

Figure \ref{fig:torus} illustrates the impact of varying each parameter on the NIR-FIR
SED. 
As $\rm R_{in}$ increases, the torus moves farther from the central engine; therefore, the emission of hotter dust grains becomes less pronounced, and the peak of the torus SED moves to longer wavelengths. The impacts of variation of the filling factor and optical depth of the clumps are intertwined. 
\reff{An increase in $\rm V_{C}$ and/or $\rm \tau_{C}$ (when the other parameters are unchanged) results in more absorption at shorter wavelengths (see the variation of a silicate emission feature to an absorption feature at 9.7 \um), and a shift of the peak of the SED to longer wavelengths due to the higher density of the clouds closer to the central engine.} 
An increase in the opacity of the homogeneous disk ($\rm \tau_{D}$) results in more scattering from the disk surface, which acts similarly to adding more dust grains with a range of temperatures (for a face-on observer). Therefore the emission from short and long wavelengths both increase. However, this will not be the same for an edge-on observer \citep[see Figures 4 and 5 in][]{Siebenmorgen2015}.
As the observer's inclination angle changes from face-on towards edge-on, the observer sees less of the emission from the inner parts of the torus (short wavelengths) and, rather than a silicate emission feature, sees an absorption feature.
Our fitting procedure, in general, does not have any prior information to limit the torus parameters, but in a few cases the observed silicate feature helped to constrain the torus model (see Section \ref{sec:quasar_sed}).

As noted in the introduction, the MIR emission from the torus comes from the absorbed and reprocessed UV and visible photons radiated by the accretion disk. Therefore, the SED shape of the primary source of emission might influence the reprocessed MIR radiation.  The primary source of radiation in \sieb\ is the \cite{RR1995} model, which describes the visible-UV continua of quasars with a broken power-law function. Instead, we use the \kd\ accretion disk model (see Section \ref{sec:AD}).  We examined the impact of various accretion disk models (including a simple blackbody, \citealt{Kubota2018} and a few others) on the torus SED and found that the SED shape longward of 1\,\um (rest-frame) is independent of the chosen accretion disk template at wavelengths shorter than 1\,$\mu$m.
i.e., the torus essentially acts as a calorimeter.

\begin{table*}[]
    \centering
        \caption{Radio properties of the quasars in our sample}
    \begin{tabular}{ l c c c c c cc }
    \hline \hline \vspace{1pt}
         Name  & Type$^{1}$ &$\rm R^{2}_{CD}$ &  Projected jets length$^{3}$& Inclination angle $^{4}$ & De-projected jets length$^{5}$  & Range of de-projected jets length $^{6}$\\
         & && (kpc) &($^{\circ}$)&(kpc)&(kpc)\\
         \hline  
\rule{0pt}{2.5ex}3C 009 & - & 0.009 & 85&52 $\pm10$ & 151&  [135,178]\\
3C 014 & --& 0.018 &213 & 42$\pm10$& 308&[261,389]\\
3C 043 & CSS & $<0.060$& 26 & $>26$& $<49$  &--\\
3C 181 & -- & 0.009 & 59&49$\pm10$ & 65&[57,77]\\
3C 186  & CSS &0.042& 16  & 42$\pm10$& 20&[17,25] \\
3C 190 & CSS & 0.098 & 25& 27$\pm10$ &124 &[94,194]\\
3C 191  & -- & 0.102& 34& 31$\pm10$& 81& [64,116]\\
3C 204 &-- & 0.087& 296& 34$\pm10$& 543  &[437,748]\\
3C 205  & -- & 0.0309 & 154 &  38$\pm10$& 251&[208,330]\\
3C 208 & -- & 0.105& 107 & 32$\pm10$& 172 &[136,243] \\
3C 212  & -- &  0.204 & 73 &  20 $\pm10$&  216 &[147,425]\\
3C 245  & -- & 1.950& 40 & 4 $_{-4}^{+10}$ & 1067 &[307,--]\\
3C 268.4 & -- & 0.091& 85 & 27$\pm10$& 205 &[155,319] \\
3C 270.1  & -- &  0.282& 103 & 16$\pm10$& 374&[235,987] \\
3C 287  & CSS &$-^{7}$& 8  &--& --   & -- \\
3C 318 & CSS & $<0.138$ & 9 &  58 $\pm10$ & 8 &[7,9]\\
3C 325 & -- & 0.003 & 124 & --&-- &--\\
4C 16.49 & -- & 0.052& 145 &39$\pm10$&  216& [180,281] \\
3C 432  & -- &  0.025& 111 &  46$\pm10$& 155&[134,190]\\
3C 454.0& CSS &  $<0.339$ & 10 & $>14$ &$<46$&-- \\
\hline
    \end{tabular}
    \begin{tablenotes}
\item[1] 1) Compact Steep Spectrum \citep[CSS, see references in ][]{Wilkes2013}
\item[2] 2) Radio core dominance ($\rm R_{CD}$, see equation \ref{eq:rcd})
\item[3] 3) Projected radio sizes, measured lobe-to-lobe from high resolution images at 5 GHz 
\item[4] 4) Jet inclination angle \citep{Marin2016}
\item[5] 5) De-projected radio sizes, measured lobe-to-lobe at 178 MHz
\item[6] 6) The range of the de-projected sizes estimated from the range of the inclination angles in column 4
\item[7] 7) We did not find the $\rm R_{CD}$ estimate for 3C 287 in the literature, so we adopted the average values of the other CSS quasars as an upper limit.
\end{tablenotes}
    \label{tab:radio}
    \end{table*}


 \subsubsection{The Radio Component} \label{sec:radio}

\reff{ 
 Table~\ref{tab:radio} lists the salient radio properties of our sample sources, including their usual classification as a compact steep spectrum \cite[CSS, see references in ][]{Wilkes2013} and their radio core dominance parameter  \citep[][]{Orr1982}, which is the ratio of 5\,GHz radio-core to extended radio-lobe emission}
\begin{equation} \label{eq:rcd}
 R_{CD} =  L_{(core, 5\,GHz)}/L_{(lobe, 5\,GHz)}    
\end{equation}
 \reff{We also list projected radio jet lengths measured lobe-to-lobe at 178~MHz (taken from the compilation of C. Willott) in Table~\ref{tab:radio}  \footnote{http://astroherzberg.org/people/chris-willott/research/3crr/} as well as the radio jets inclination angle, within $\pm10^{\circ}$ \citep[estimated from the radio core fraction,][]{Marin2016}. 
 The de-projected radio jet lengths (and their range) are also listed. }

To fit the radio emission from the quasars in our sample, we started by considering a relativistic electron population with a power-law energy distribution in a magnetic field of strength $B$:

\begin{equation}\label{eq:edensity}
 N(E)\propto E^{-p}
\end{equation}
Synchrotron emission is generated by these relativistic electrons spiraling around magnetic field lines. \reff{While some of 
this emission is absorbed by the electrons in optically thick regions (synchrotron self-absorption), radio photons from optically thin regions reach the observer.} We can formulate synchrotron emission transitioning  from an optically thick, self-absorbed region to an optically thin region with the radiative transfer equation \citep[for more details, see Section 4.4--4.7 in] []{GH2013}:

\begin{equation}\label{eq:radiative_transfer}
I(\nu) = \frac{\epsilon_{\nu}}{\kappa_{\nu}} \ (1- e^{-\tau_{\nu}}); \;\;\;\; \tau_{\nu} \equiv R\kappa_{\nu}
\end{equation}

in which $\rm \tau_{\nu}$ is the spectral optical depth, $\rm R$ is the size of the emitting region,  $\epsilon_{\nu}$ is the emissivity, and $\kappa_{\nu}$ is the specific absorption coefficient. \reff{These can be approximated by}

\begin{equation}\label{eq:epsilon}
\epsilon_{\nu} \propto B^{(p+1)/2} \ \nu^{-(p-1)/2}
\end{equation}

\begin{equation} \label{eq:kappa}
\kappa_{\nu} \propto B^{(p+2)/2} \ \nu^{-(p+4)/2}.
\end{equation}{}

 When $\tau_{\nu} \gg1$ (in a self-absorbed regime), the second component in the radiative transfer equation becomes negligible, and equation \ref{eq:radiative_transfer} can be simplified to: 
 
\begin{equation}
I(\nu) =I_{0}(\nu) \propto B^{-1/2} \ \nu^{5/2}
\end{equation}

\reff{which is independent of $\rm p$ \citep[see Section 4.5 in] []{GH2013}.} Setting $\nu_{t}$ to be the frequency at which the transition from optically thick to thin  occurs, then 
\begin{equation}
\tau_{\nu_{t}} \equiv R\kappa_{\nu_{t}}=1
\end{equation}

and using equation \ref{eq:kappa}, we then obtain 

\begin{equation}
\nu_{t} \propto [RB^{(p+2)/2}]^{2/(p+4)}
\end{equation}

Therefore

\begin{equation}
\tau_{\nu} \equiv R\kappa_{\nu} = (\frac{\nu_{t}}{\nu}) ^{(p+4)/2} = (\frac{\nu_{t}}{\nu}) ^{(p-1)/2 + 5/2}
\end{equation}

Now if  $\alpha_{2} = -(p-1)/2$ is the spectral index in the optically thin region, and $\alpha_{1}=5/2$ is the spectral index in the optically thick region, we can rewrite equation \ref{eq:radiative_transfer} as

\begin{equation} 
I(\nu) \propto (\frac{\nu}{\nu_{t}})^{\alpha_{1}} \ [1- \exp(-(\frac{\nu_{t}}{\nu})^{\alpha_{1}-\alpha_{2}})]    
\end{equation}

As the electron population ages, it loses its energy, and eventually, the synchrotron radiation terminates at a cutoff frequency. This can be approximated by adding an
exponential factor \citep{Polletta2000}:  

\begin{equation} \label{eq:polletta}
L_{\nu} \propto (\frac{\nu}{\nu_{t}})^{\alpha_{1}} \ [1- \exp(-(\frac{\nu_{t}}{\nu})^{\alpha_{1}-\alpha_{2}})]  \ e^{-\frac{\nu}{\nu_{\rm cutoff}}}
\end{equation}

 Equation \ref{eq:polletta} describes radiation from both optically thick and thin regions, which terminates at high energies due to energy loss.
This equation is based on many assumptions, including the charged particle being an electron, the electron population's energy having a power-law distribution,
 the source is homogeneous, and therefore $\alpha_{1}$ =2.5.  In order to describe the radio emission from the quasars in our sample, we used the general form presented in equation~\ref{eq:polletta} and modified it based on different conditions. 

{\sl Model 1--Single Power-Law with an Exponential Cutoff:} In 
an AGN in which the radio emission is dominated by radiation from the optically thin lobes, a power-law can successfully describe the radio emission \citep{Polletta2000}:

\begin{equation} \label{eq:powerlaw}
L_{\nu} \propto (\frac{\nu}{\nu_{\rm cutoff}})^{\alpha_{2}}  \ e^{-\frac{\nu}{\nu_{\rm cutoff}}}
\end{equation}

{\sl Model 2--Double Power-Law with an Exponential Cutoff:}
The added presence of compact structures such as radio cores and hot spots may cause the spectral shape to deviate from a single power law. In this model, we assumed a double power-law similar to equation~\ref{eq:polletta}. \reff{However, $\alpha_{1}$ was treated as a free parameter, not fixed at 2.5 (because we do not expect a single, homogeneous component), and $\nu_{t}$ is the transition frequency from $\alpha_{1}$ to $\alpha_{2}$-dominated regions.} 

{\sl Model 3--Parabola with a Cutoff:}
Relaxing the homogeneous assumption, our third model assumes a parabolic function that approximates a superposition of multiple power-law components, some optically thick and some thin \citep{Cleary2007}:

\begin{equation} \label{eq:cleary}
\log L_{\nu} \propto -\beta (\log \nu- \log \nu_{t})^{2}  + \log (e^{-\frac{\nu}{\nu_{\rm cutoff}}})
\end{equation}

$\beta$ in this equation indicates the curvature of the parabola.

Overall, equations~\ref{eq:polletta} or \ref{eq:cleary} can successfully describe the emission from a source with a mix of optically thick and thin regions (e.g., a lobe-dominated quasar with bright hot spots or a moderately bright core). 

{\sl Model 4--Parabola with a Cutoff \& Double Power-Law with an Exponential Cutoff: }
In quasars with bright cores, the superposed self-absorbed components may generate spectral shapes not amenable to fitting with simple models. Our fourth model is a combination of equations ~\ref{eq:polletta} and \ref{eq:cleary}, where equation ~\ref{eq:polletta} is solely used to describe the radio core emission, and equation ~\ref{eq:cleary} describes radiation from the lobes and hot spots. In this case, the core and lobe components each have their own cutoff frequency.

We fit the above four models to the observed radio photometry using the \texttt{MPFIT} nonlinear least-square fitting function in IDL \citep{Markwardt2009}. 
The spectral slopes $\alpha_{1}$ and $\alpha_{2}$ in each model as well as $\beta$ and $\nu_{t}$ are free parameters, and the cutoff frequency was limited to $10^{10}<\rm \nu_{cutoff}<10^{14}$\,Hz.

\subsection{The Host Galaxy Component} \label{sec:magphys}

We used the \m SED code \citep{dc2008,dcr2015} to account for the emission from the host galaxy. 
\m is capable of accounting simultaneously for different levels of star formation activity, stellar populations, dust obscuration, and star formation histories for galaxies at different redshifts.  \reff{We note that \m does not include any AGN component, and to fit the emission from  AGN,  we used the models described in Section \ref{sec:agn}.}

\m is built upon the energy balance technique, which links the UV and visible emission from the young stellar population to the IR emission from dust. In other words, starlight is the only source of dust heating, and the energy absorbed by dust is equal to the re-radiated energy. The stellar emission from UV to NIR wavelengths is modeled with the \cite{bc2003} spectral population synthesis model \citep[assuming a][IMF]{chabrier2003} attenuated by dust following \cite{charlot2000simple}.  The model assumes the young stars form in dense clouds (i.e., giant molecular clouds); when younger, their emission is attenuated by the dust in their birth cloud and the ambient ISM, but as they age, the birth clouds disappear on a time scale of $10^7$\, yr, and the stellar emission is then absorbed only by the diffuse ISM \citep{charlot2000simple}. \m considers four dust components, each including grains with a different size and temperature: polycyclic aromatic hydrocarbons (PAHs) grains, hot grains with temperatures in the range 130--250\,K, grains in thermal equilibrium with the temperature of 30--60\,K, and cold dust grains with adjustable equilibrium temperature in the range 15--25\,K.  The stellar birth clouds contain the first three dust components, and cold dust only exists in the ambient ISM (see \citealt{dc2008} for more details).

Although the original \dc\ \m code covers a broad wavelength range from 912\,\AA\ 
to 1\,mm, here we use the updated \dcr\ version, which includes the galaxy contribution at radio wavelengths \citep[for more details see Section 3.2 in][]{dcr2015}.  In addition to broader wavelength coverage, \reff{the more recent version} includes a continuous delayed exponential SFH:
\begin{equation}
    \Psi (t) \propto \gamma^{2} t \exp(-\gamma t)
\end{equation}
in which t is the time since the onset of star formation and $\rm \gamma=1/\tau_{SF}$ is the inverse of the star formation time scale \citep[see][]{Lee2010,dcr2015}. 
\dcr\ also included absorption by the intergalactic medium (IGM), which can strongly impact the UV emission in 
high-redshift galaxies such as the ones treated here.

Although the \dcr\ model includes galaxy emission at radio wavelengths, the radio emission (both thermal and non-thermal components) in galaxies is 4--5 orders of magnitude lower than that in radio-loud AGN. Therefore AGN are the dominant sources of radio emission in the
present analysis.


\subsection{ Dereddening and Absorption Corrections} \label{sec:dered}

In order to obtain the intrinsic SED of the AGN in our sample, it is necessary to correct the photometry for the absorption occurring at various wavelengths. In brief, these corrections include correcting the X-ray observations for host galaxy and Milky Way absorption, correcting the UV--NIR photometry for Milky Way absorption, and correcting the visible-UV radiation from the accretion disk for dusty torus absorption.

X-ray photons are absorbed by the gas in their host galaxies (i.e., intrinsic) and along the line of sight in the Milky Way. We adopted the best estimates of the X-ray luminosity from \wilkes, i.e., corrected for both intrinsic and Galactic absorption. Additionally, we corrected the UV--NIR (0.91--13.0 $\mu$m) photometry for absorption by the Milky Way using the attenuation law of $\tau_{\lambda} \propto \lambda^{-0.7}$ from \cite{charlot2000simple}.

To estimate the extinction of the accretion disk emission from the obscuring structure in the torus, we used the method of \citeauthor{Siebenmorgen2015} \citep[2015, see also][]{krugel2009}. According to this method, the effective optical depth for any templates (with any combination of dust clouds and homogeneous disk) can be obtained via comparison to the flux of that template in the absence of dust \citep[see Section 2.7 in][]{Siebenmorgen2015}: 
\reff{
\begin{equation}\label{eq:torus_corr}
\tau_{\rm eff} = - \ln \frac{f_{(best-fit -template)}}{f_{(no-dust-template)}}
\end{equation}}

 Thus, we first determined the template which best fits the NIR--FIR data corresponding to the AGN emission (not the host galaxy, see Section \ref{sec:fitting} for more details) and then applied equation \ref{eq:torus_corr} to 
estimate the effective optical depth. The visible--UV  photometry (corresponding to the accretion disk emission) was then corrected by a factor of $e^{\tau_{\rm eff}}$ for the absorption in the torus.
Depending on the best fit, ${\tau_{\rm eff}}$ may be a negative or positive number indicating scattering or absorption by the dust structure. An edge-on observer is not able to see the scattered light from the torus and is mainly affected by absorption, i.e., a positive  ${\tau_{\rm eff}}$, while a face-on observer sees the scattered light from the torus and is not affected by absorption, i.e.,  a negative ${\tau_{\rm eff}}$  \citep[see][]{Siebenmorgen2015}.

\subsection{Fitting Methodology} \label{sec:fitting}

In this Section, we describe our fitting methodology and the steps in which the various attenuation and absorption corrections from Section \ref{sec:dered} are applied.  

In radio-loud quasars, AGN emission dominates that of the host galaxy over most of the SED. Therefore, we allowed the contribution of the AGN to the photometry at submm to UV wavelengths to vary from 95\% to 65\% (in bins of 5\%) of the total.  At radio and X-ray wavelengths, the host galaxy contribution is orders of magnitude smaller than the AGN. 
At IR to UV wavelengths, torus and accretion disk emission, which dominate the SED, were allowed to vary independently from one another. \reff{The variation of the AGN components contributions results in an iterative process:}

{\sl Step 1:} The fitting procedure starts by determining the torus templates which best fit the photometry associated with the AGN at IR wavelengths. The torus templates of Siebenmorgen et al. are normalized to an AGN of luminosity $\rm 10^{11}\,L_{\odot}$ (Section \ref{sec:torus}). \reff{Therefore, we integrate the luminosity of each AGN within the range of 2--45\,\um (rest-frame)  and normalize the torus templates according to this integrated luminosity for each source.} We then determine the best-fit torus model using $\chi^{2}$ minimization.

{\sl Step 2:} We corrected the visible--UV photometry associated with the accretion disk for the torus reddening, 
implementing the correction factor obtained from equation \ref{eq:torus_corr}.
To do this, we assumed the line-of-sight to the accretion disk is the same as that for the torus.
\reff{We then fitted the accretion disk templates, built for this sample (see Section \ref{sec:AD}), to the visible-UV and X-ray photometry and identified the best-fit accretion disk template via $\chi^{2}$ minimization.}

{\sl Step 3:} We fitted the radio photometry with the four models described in Section \ref{sec:radio} and determined the best fit using  $\chi^{2}$ minimization. 
We normalized the radio models to the 5\, GHz photometry. In the fourth radio model, we used the 5\, GHz core flux density to normalize the core component and the remaining flux density to normalize the parabola component. We inspected the radio images in several bands (178 MHz, 5 GHz, 8 GHz, and 15 GHz, if available) to
check the consistency of any preference for  multi-component models with different spectral indices with the presence of bright hot spots and/or core emission.

{\sl Step 4:} We subtracted the best-fit AGN component (radio, torus, and accretion disk) from the total observed photometry and fit this residual with MAGPHYS to account for the host galaxy contribution. To include the upper limits in MAGPHYS, we followed the prescription of \dcr\ in which the flux densities are set to zero, and the upper limit values are set as the uncertainty.

{\sl Step 5:} We found the total fit by combining the AGN and the host galaxy components. 

{\sl Step 6:}
\reff{We repeated Steps 1--5 above for all combinations of the torus and the accretion disk normalizations (95\%-65\% in bins of 5\%). We also examined the cases with 100\% or $<65\%$ contributions from the AGN components. However, none of those resulted in a better fit than the one presented. In particular, the torus component can not replicate the emission at FIR-submm wavelengths, and an underlying host galaxy component is required. We identified the best fit as the one that not only results in low $\chi^{2}$ values for the total fit but individual components.}

We note that after identifying the best-fit torus model via $\chi^{2}$ minimization, we examined the impact of varying the torus parameters on the visible-UV SED and the total fit. This is particularly important for the so-called red quasars in our sample. These are reddened Type 1 AGN in which the visible-UV SED lacks the big blue bump
and have red MIR colors 
\citep[e.g.,][]{Benn1998,Cutri2001,Lacy2004,Georgakakis2009,Kim2018}. 
\reff{ In our fitting routine, we assume that the reddening is due to the torus obscuration and examined the dependence of the shape of the intrinsic visible-UV SED on the torus parameters to obtain a reasonable UV bump in the red quasars. In this procedure, the obscuration from the host galaxy at visible-UV bands is assumed to be negligible. For the red quasars in our sample (3C~14/190/212/325), the galactic hydrogen column densities $\rm N_{H,gal}$ are 15--375 times less than the torus $\rm N_{H}$ \citep{Wilkes2013}, which indicates the absorption from the host galaxy is much less than the torus \footnote{\reff{3C~68.1, excluded from our sample (see Section \ref{sec:data}), is a red quasar that required additional correction from the host galaxy}}.} We describe the best-fit model to each of these sources in Section \ref{sec:fitting_analysis}; however, understanding the nature of red quasars is beyond the scope of this paper.

\begin{table*}[] 
    \centering
        \caption{Parameters of the best-fit torus and accretion disk model}
    \begin{tabular}{ l c c c c c c c c c}
    \hline \hline \vspace{1pt}
\rule{0pt}{2.5ex}Name  & $\rm R_{in}(pc)^{1}$& $\rm V_{C}(\%)^{2}$& $\rm \tau_{C}^{3}$&$\rm \tau_{D}^{4}$ &$\rm \theta_{T}(^{\circ})^{5}$ & $\rm \theta_{AD}(^{\circ})^{6,9}$&$\rm M_{BH}(\times 10^{9} M_{\odot})^{9}$ &$\rm \log (\lambda_{Edd})^{7,9}$& $Spin^{8,9}$
\\ 
\hline 
\rule{0pt}{2.5ex}3C 009 & 3.74 & 1.5 &  0&1000&$52$& 48$\pm6$&3.8$\pm1.2$&$-0.4\pm0.2$&0.98$\pm0.04$  \\

3C 014 & 4.53 & 77.7 &  0&100&67& 57$\pm7$&9.6$\pm$1.4&$-0.4\pm0.4$&0.92$\pm0.02$  \\
3C 043 & 1.80 & 1.5 &  45&100&60& 70$\pm2$&2.1$\pm$0.7&$-1.0\pm0.2$&0.99$-0.01$   \\
3C 181 & 2.85 & 1.5 &  45&300&52& 59$\pm8$&1.7$\pm$0.7&$-0.6\pm0.1$&0.92$\pm0.08$   \\
3C 186  & 2.75 & 1.5 &  4.5&300&52& 62$\pm7$&1.5$\pm$0.5&$-0.6\pm0.5$&0.99$-0.02$   \\
3C 190 & 5.95 & 1.5 &  45&300&60& 63$\pm4$&6.0$\pm$1.3&$-1.0\pm0.1$&0.99$-0.03$  \\
3C 191  & 6.53 & 1.5 &  45&300&43& 43$\pm5$&1.8$\pm$0.0&$-0.6\pm0.2$&0.92 $\pm0.02$  \\
3C 204 & 2.59 & 77.7 &  0&300&52& 55$\pm6$&2.9$\pm$0.0&$-1.0\pm0.2$&0.98 $\pm0.03$  \\
3C 205 & 5.79 & 7.7 &  4.5&300&52& 55$\pm6$&9.8$\pm$0.0&$-0.8\pm0.5$&0.92  $\pm0.01$  \\
3C 208 & 2.09 & 1.5 &  4.5&300&52& 48$\pm6$&1.5$\pm$1.2&$-0.4\pm0.5$&0.99 $-0.01$   \\
3C 212 & 1.71 & 7.7 &  4.5&300&60& 53$\pm2$&6.0$\pm$0.0&$-1.2\pm0.1$&0.96  $\pm0.00$ \\
3C 245 & 7.94& 1.5 &  0&30&19& 26$\pm7$&1.1$\pm$0.0&$-0.6\pm0.2$&0.98 $\pm0.02$  \\
3C 268.4 & 2.29 & 1.5 &  45&1000&52& 52$\pm7$&6.2$\pm$2.3&$-1.0\pm0.2$&0.99  $-0.04$ \\
3C 270.1  & 2.55& 1.5 &  45&1000&52& 50$\pm5$&4.6$\pm$0.9&$-1.0\pm0.9$&0.98 $\pm0.02$  \\
3C 287  & 1.16 & 38.5 &  0&1000&19& 26$\pm0$&1.5$\pm$0.8&$-1.2\pm0.1$&0.99  $\pm0.00$  \\
3C 318 & 1.30 & 1.5 &  45&1000&33& 38$\pm9$&2.4$\pm$0.8&$-1.6\pm0.2$&0.98 $\pm0.02$  
\\
3C 325 & 0.75 & 1.5 &  13.5&300&60& 49$\pm2$&3.8$\pm$0.0&$-1.4\pm0.0$&0.88 $\pm0.02$   
\\
4C 16.49 & 1.19 & 7.7 &  0&1000&43& 55$\pm0$&2.1$\pm$0.7&$-1.4\pm0.0$&0.94  $\pm0.02$
\\
3C 432 & 3.07 & 38.5 &  0&1000&33& 46$\pm9$&4.7$\pm$1.4&$-1.2\pm0.2$&0.99$-0.01$ 
\\
3C 454.0& 4.53 & 1.5 &  4.5&1000&43& 51$\pm9$&0.8$\pm$0.0&$\;\;\;0.0\pm0.1$&0.99$-0.04$
\\
\hline
    \end{tabular}
    \begin{tablenotes}
\item[1] 1) The inner radius of the best-fit torus ($\rm R_{in}\propto \sqrt{\rm L_{AGN}/10^{11} L_{\odot}}$)
\item[2] 2) The volume filling factor of the clumps
\item[3] 3) The optical depth of the individual clumps
\item[4] 4) The optical depth of the homogeneous disk midplane
\item[5] 5) The inclination angle of the torus 
\item[6] 6) The inclination angle of the accretion disk
\item[7] 7) The Eddington ratio (  $\lambda_{Edd} \propto$  $\rm L_{bol}/\rm L_{Edd}$) 
\item[8] 8) The dimensionless spin parameter $\rm a\equiv Jc/GM_{BH}^{2}$ where $\rm J$ is the angular momentum of the SMBH
\item[9] 9) \reff{\maz{The reported uncertainties in accretion disk parameters are standard deviations derived from the 5\% of the fits with the lowest $\chi^2$ values }}
   \end{tablenotes}
   \label{tab:best_fit_torus}
    \end{table*}

\smallskip\noindent

\begin{table*}[] 
    \centering \caption{The parameters of the best-fit radio model}
    \begin{tabular}{lcccccccc}
    \hline \hline
         ID& Best radio model$^{1} $&$\rm \nu_{t}$&$\rm \nu_{t,jet}$&$\alpha_{1}$&$\alpha_{2}$ &$\beta$&$\rm \nu_{cutoff}$&$\rm \nu_{cutoff,jet}$ \\
         \hline
3C 009    &2&5.5e+07& - &$-$0.09&$-$1.04& - &5.7e+11 & - \\
3C 014   &2&3.8e+10& - &$-$0.91&$-$2.15& - &1.0e+13 & - \\
3C 043 &1& - & - & - &$-$0.75& - &1.5e+12 & - \\
3C 181&1& - & - & - &$-$0.94& - &8.1e+11 & - \\
3C 186  &3&5562.3& - & - & - &10.28&8.3e+12 & - \\
3C 190  &2&4.7e+07& - &1.98&$-$0.92& - &4.9e+12 & - \\
3C 191&1& - & - & - &$-$0.99& - &3.0e+12 & - \\
3C 204  &2&1.1e+07& - &1.37&$-$1.10& - &1.0e+12 & - \\
3C 205  &2&5.1e+07& - &1.34&$-$0.98& - &2.1e+11 & - \\
3C 208  &2&1.9e+08& - &$-$0.14&$-$1.16& - &1.0e+12 & - \\
3C 212&4&1000.1&1.5e+10&2.20&$-$0.52&7.46&5.0e+11&7.5e+11\\
3C 245&4&3814.4&7.9e+08&2.50&$-$0.35&9.41&8.1e+12&5.1e+11\\
3C 268.4&2&6.9e+09& - &$-$0.71&$-$1.17& - &3.0e+12 & - \\
3C 270.1&2&5.6e+07& - &1.35&$-$0.90& - &1.0e+12 & - \\
3C 287  &2&1.8e+09& - &$-$0.13&$-$0.79& - &1.3e+12 & - \\
3C 318  &2&4.3e+08& - &$-$0.07&$-$1.01& - &2.0e+11 & - \\
3C 325  &2&1.7e+09& - &$-$0.50&$-$1.12& - &2.2e+11 & - \\
4C 16.49&2&4.2e+08& - &$-$0.40&$-$1.17& - &1.0e+11 & - \\
3C 432  &2&1.1e+09& - &$-$0.64&$-$1.19& - &2.8e+11 & - \\
3C 454.0&3&600.4& - & - & - &5.92&1.8e+12 & - \\

\hline
    \end{tabular}
    \label{tab:best_radio_fit}
       \begin{tablenotes}
\item[1] 1) \reff{1: single power-law with an exponential cutoff, 2: single power-law with an exponential cutoff, 3: parabola with a cutoff, 4: parabola with a cutoff \& double power-
law with an exponential cutoff}
\end{tablenotes}
\end{table*}

\begin{figure*}[h!t]
\includegraphics[width=0.93\textwidth, angle =90]{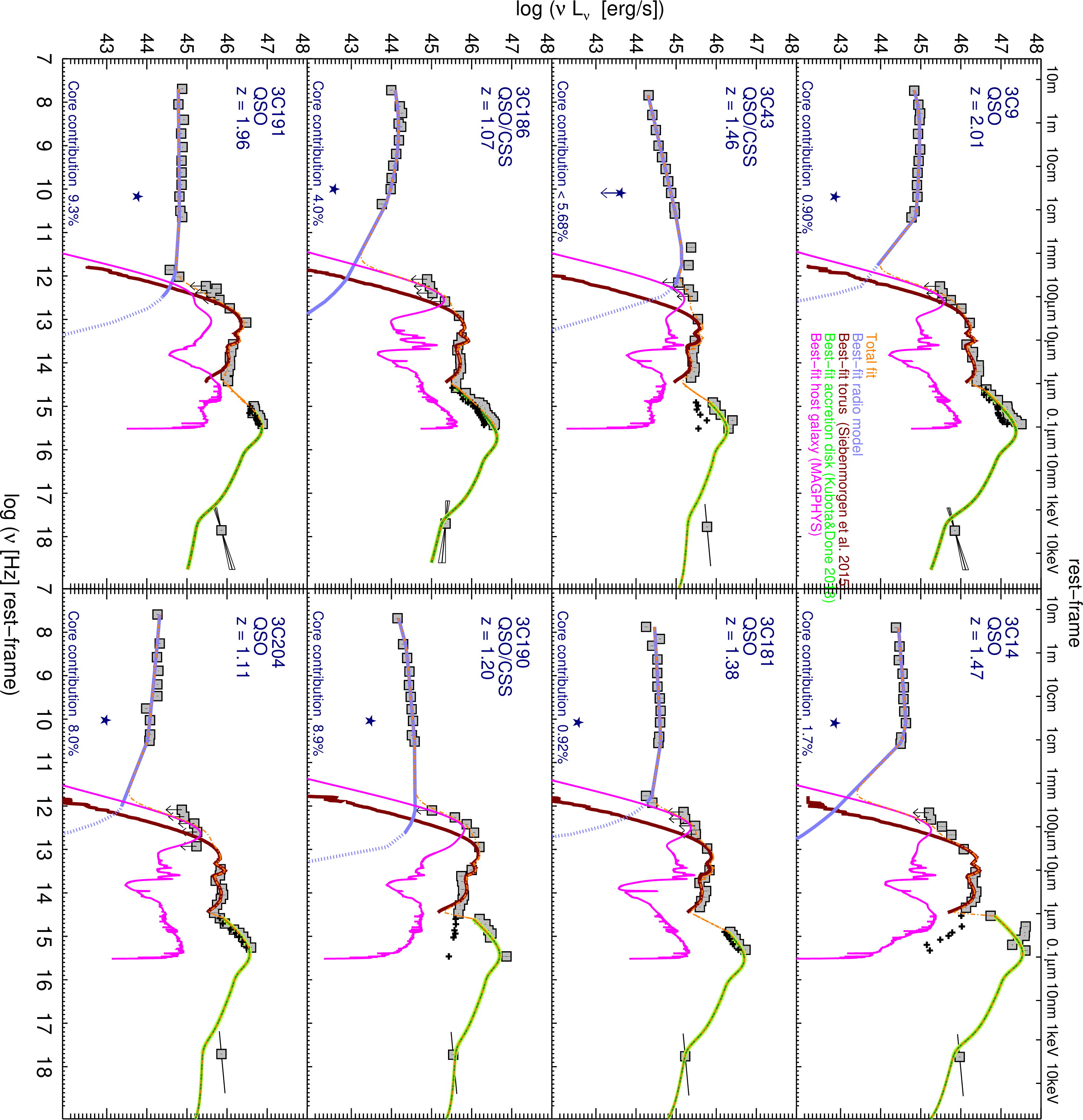} 
    \caption{\reff{The full suite of 
photometry with the best-fit models of the 3CRR quasars at \z. The gray data points show the absorption-corrected broadband photometry, and the black plus signs in visible-UV bands are the photometric points before the torus obscuration correction. The lines represent radio emission from the core, jets, hot spots, and lobes (light blue), infrared emission from the torus (dark red), thermal visible to X-ray emission from the accretion disk (green), the underlying host galaxy emission (magenta), and the total fit (orange). The radio emission truncates at the cutoff frequency (where the dotted line begins). The dark blue star indicates the core contribution to the 5 GHz photometry estimated from $\rm R_{CD}$. The black line (and the bow tie for sources with higher X-ray counts) at X-ray indicates the power-law fit to the absorption corrected data.}}
\label{fig:sed}
\end{figure*}

\renewcommand{\thefigure}{\arabic{figure} (Cont.)}
\addtocounter{figure}{-1}

\begin{figure*}[h!t]
\includegraphics[width=0.92\textwidth, angle =90]{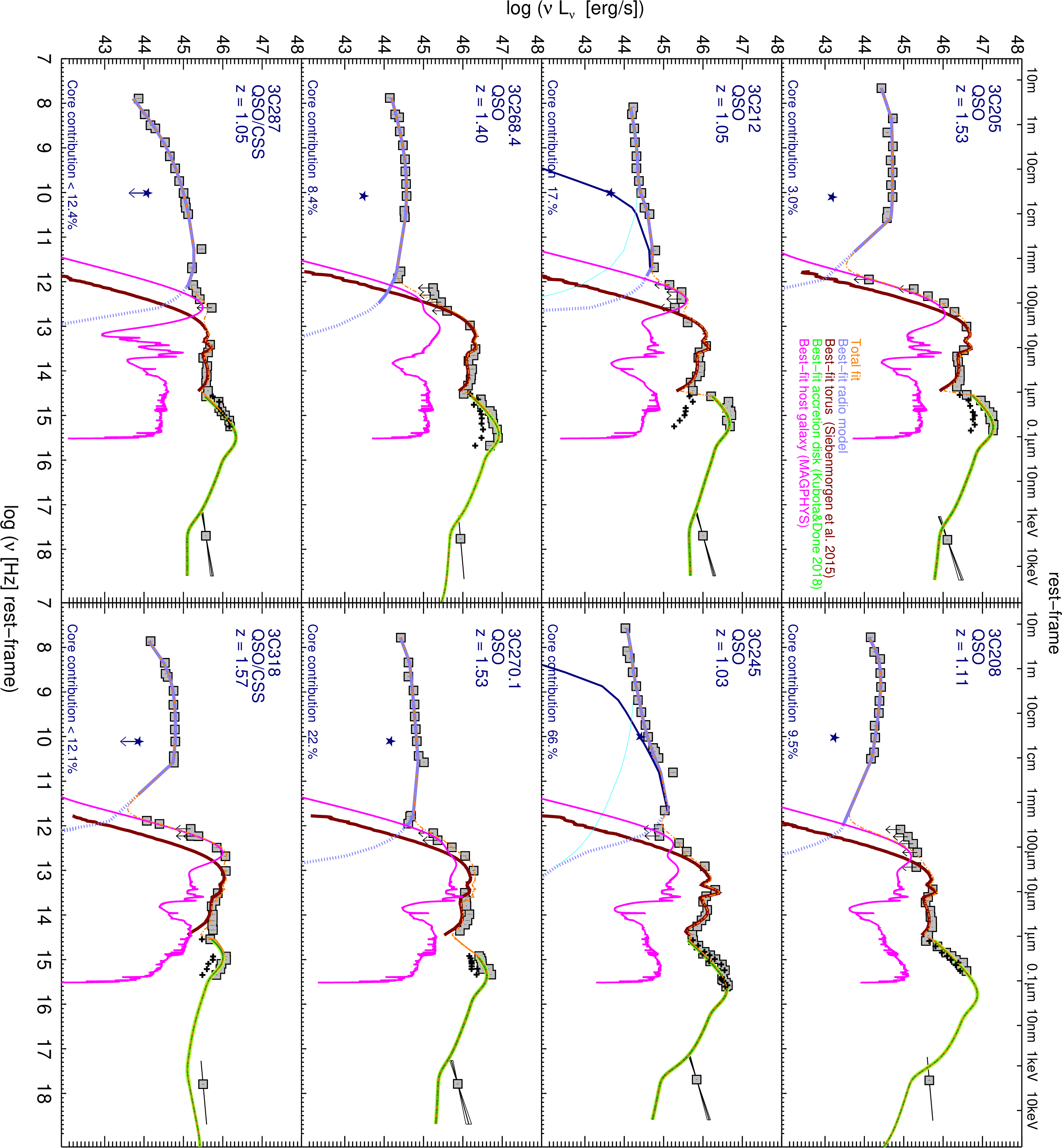} 
   \\
         \caption{\reff{In the case of 3C 212/245, the cyan parabola component indicates the emission from the extended structures, while the dark blue component indicates the emission from the core.}}
\end{figure*}

\renewcommand{\thefigure}{\arabic{figure}}

\renewcommand{\thefigure}{\arabic{figure} (Cont.)}
\addtocounter{figure}{-1}
\begin{figure*}[h!t]
   \includegraphics[width=0.46\textwidth , angle=90]{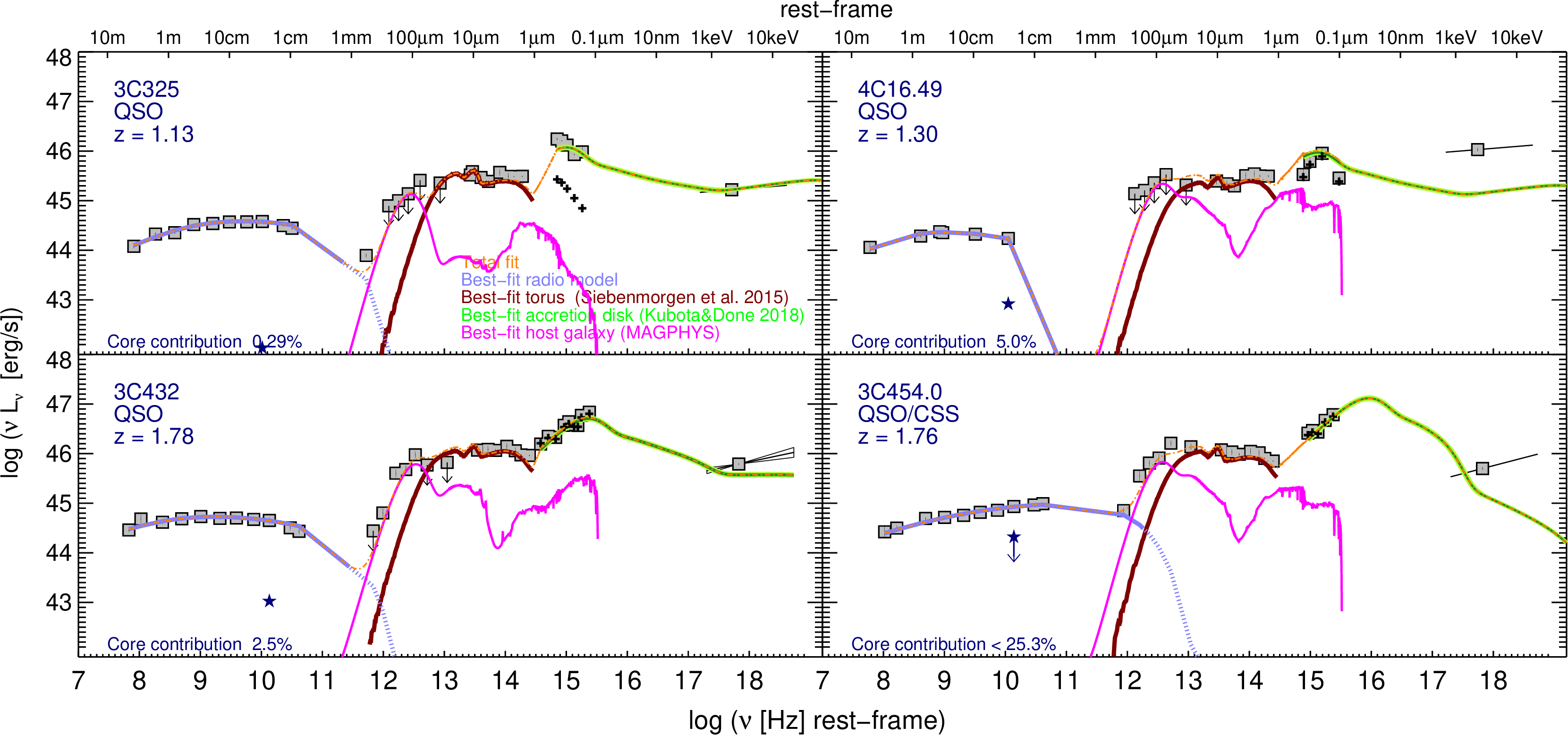} 
             \caption{}
\end{figure*}
\renewcommand{\thefigure}{\arabic{figure}}

\section{Radio to X-ray SED ANALYSIS OF THE 3CRR QUASARS at \z} 
\label{sec:fitting_analysis}
\reff{In this section, we apply ARXSED to the 3CRR quasar sample at \z, and describe the results.}
We first describe the details of the best-fit AGN model for individual sources and then summarize the commonalities among sources.

\subsection{Fitting Results for Individual Sources} \label{sec:quasar_sed}

The parameters for the best-fit accretion disk and torus and radio models are given in Tables \ref{tab:best_fit_torus} and  \ref{tab:best_radio_fit}. 
Figure \ref{fig:sed} shows the full suite of 
photometry together with the best-fit models. The gray data points are the absorption-corrected broadband photometry, and the black plus signs in visible-UV bands indicate the photometric points before the torus obscuration correction. The components in these plots are radio component from the core, jets, and lobes (light blue), infrared emission from the torus that contributes at rest-frame wavelengths longward 1 \um (dark red), the accretion disk component that accounts for the thermal optical to X-ray emission (green), the host galaxy component obtained from MAGPHYS SED fitting code (magenta) and the total fit (orange). The dark blue star indicates the core contribution to the 5 GHz photometry estimated from the $\rm R_{CD}$ presented in Table \ref{tab:radio}.
\reff{The X-ray photometry was extracted within a 2.2\arcsec\ aperture centered on the X-ray source position, sized to match the full Chandra point-spread function, and so is dominated by the nuclear X-ray emission \citep{Wilkes2013}. The black line (and the bow tie for sources with higher X-ray counts) at the X-ray point indicates the power-law fit to the absorption corrected data.} The SED fits are presented in 3CRR number order below:


\textbf{3C 9} has deprojected jets of $\sim151$ kpc and a core contribution of less than $1\%$ at 5 GHz. The radio emission in this source is best described as a broken power-law (with an exponential cutoff at 570 GHz, see Table \ref{tab:best_radio_fit}), which is consistent with the presence of multiple features \citep[i.e., jet, counter-jet, bright hot spots based on 4.9 GHz VLA data,][]{Bridle1994}.

The emission from the torus is best described with a combination of clumps with negligible opacity and a homogeneous disk with the highest allowed optical depth (see Table \ref{tab:sieb_par}). The filling factor and optical depth of the clumps indicate that there is not much obscuration from the dust clouds along the line of sight. Therefore, the correction of the visible-UV emission from the torus obscuration is negligible. For this object, the inclination angle from the best-fit torus model is $52^{\circ}$  while
from the best-fit accretion disk template, it is $63^{\circ}$. As noted in Section \ref{sec:AD} the accretion disk templates are built within the range of $\pm12^{\circ}$ from the inclination angle of the best-fit torus model.

The accretion disk template lies below the X-ray data, suggesting the presence of a non-thermal X-ray component. \textit{Chandra} observations \citep{Fabian2003}
found extended X-ray emission on both sides of the nucleus of 3C 9 coincident with the radio structure and suggested it is probably due to non-thermal inverse Compton emission from the interaction of the relativistic electrons with the cosmic microwave background. 
\reff{While the X-ray data used in our analysis is nuclear, there might still be some contribution from the extended X-ray emission, which might result in the discrepancy between the accretion disk fit and X-ray data.}

 
\textbf{3C 14} is a red quasar \citep{Smith1980} with less than 2\% core contribution at 5 GHz and  extended radio jets of 308 kpc. The radio emission is best described with a broken power-law with an exponential cutoff at
FIR wavelengths, consistent with the multiple radio structures present in MERLIN 18 cm data \citep{Akujor1994}. SED fitting shows that the non-thermal emission at submm/FIR wavelengths is negligible.  

The emission from the torus is best described with a combination of clumps with negligible opacity and a homogeneous disk with moderate optical depth. \reff{However, this, combined with the inclination angle of $67^{\circ}$ results in substantial(more than one order of magnitude) obscuration of the visible-UV emission.}
The visible-UV SED before obscuration correction (shown in black plus signs) is consistent with its classification as a red quasar. As noted in Section \ref{sec:fitting}, we examine the impact of the variation of the torus parameters on the intrinsic visible-UV SED in our fitting procedure to make sure the best-fit torus model results in a reasonable accretion disk SED. Our SED fitting indicates that the classification of 3C 14 as a red quasar is related to the inclination angle rather than a dustier torus. Using the length of the radio jets as a proxy of AGN maturity \citep[e.g.,][]{pece15} indicates that 3C 14 is a mature quasar. Therefore it is unlikely for this source to be in an evolutionary phase in which it turns from an obscured quasar into an unobscured one. \reff{To investigate the impact of the host galaxy obscuration on the visible-UV SED of this source, we will include this source when analyzing the NLRG in our sample and examine whether the correction from the host galaxy can improve the SED fit at visible-UV bands.}

The X-ray spectrum is well fit by the accretion disk model, implying no significant contribution from the radio structures at X-rays.


\textbf{3C 43} is a CSS quasar with jets of $<49$ kpc and a core contribution of $< 6\%$ at 5 GHz \citep{Akujor1991,Ludke1998}. The radio emission is best described with a single power law with an exponential cutoff at FIR wavelengths. \cite{Akujor1991} indicate that this source has a misaligned and asymmetric structure with a sharp bent jet suggesting that the presence of the bent structure in this source could be due to the interaction of the jets with the ISM.

The best-fit torus model is a combination of clumps with a small volume filling factor and a homogeneous disk with low opacity that result in a small amount of obscuration of the visible-UV emission.

As illustrated in Figure \ref{fig:sed} the \reff{mm and submm} data are dominated by non-thermal radiation from the radio structures. \reff{The submm-excess could be attributed to the interaction of the jets  with the gas along their paths. Further investigation with ALMA is required to understand the nature of submm radiation. However, the best-fit accretion disk template also confirms the presence of a significant possibly jet-triggered X-ray component. \cite{pece15} reported the presence of a bright nearby object dominating at PACS 160 \um band. The potential non-thermal emission from the radio structures and the contaminating nearby object add uncertainty to the estimated host galaxy SFR.} 


\textbf{3C 181} has radio jets of 65 kpc with $<$1\% core contribution at 5 GHz \citep{Mantovani1992}. The radio emission is best described with a single power-law with an exponential cutoff at 814 GHz. \cite{Willott2002} investigated the 850\um band SCUBA observation of this source and predicted no non-thermal component contributing to this wavelength.

The best-fit torus is a combination of clumps with a small filling factor and a homogeneous disk with moderate opacity, which results in little correction of the visible-UV emission. 

Comparing the best-fit accretion disk with the X-ray data does not indicate the presence of an underlying X-ray component triggered by the radio structures.

\textbf{3C 186}
 is a very well studied CSS \citep{ODea1998} with radio jets of 20 kpc and 4\% core contribution at 5 GHz \citep{Spencer1991,Ludke1998}. Radio emission is best described with a parabola (a combination of multiple power-laws). MERLIN observation at 1.6 MHz indicates a one-sided jet and two bent lobes at 60$^{\circ}$ and 90$^{\circ}$ with respect to the source axis, creating a S-shaped source \citep{Spencer1991, Ludke1998}. The curvature in the lobes is presumably due to the sufficiently dense ISM that affects the expansion of the radio structure.

The best-fit torus is a combination of clumps with a small filling factor and a homogeneous disk with moderate opacity, which results in no significant correction of the visible-UV emission.

The \textit{Chandra} and \textit{Hubble Space Telescope} (\textit{HST}) observations indicate that 3C 186 lives in an overdense region, which is most likely a cluster of galaxies \citep{Hilbert2016,Siemiginowska2010}. The \textit{HST} images indicate a blob of star formation activity $2"$ (corresponding to $\sim$ 16 kpc at $z\sim1$) away from the central engine, perpendicular to the direction of the jets \citep[see Figure 10 in ][]{Hilbert2016}. \footnote{\reff{We note that the \textit{HST} data are not included in our SED analysis.}}
This blob could be either from the star formation activity of other members of the cluster or from the host galaxy itself. In the  latter case, given that the jets' direction is perpendicular to the blob, the star formation activity is not jet-related. Comparing the best-fit accretion disk with the X-ray data does not indicate the presence of an underlying jet-triggered X-ray component.


\textbf{3C 190} is a red quasar \citep{Ishwara2003} CSS and has de-projected radio jet lengths of 124\,kpc, and less than 1\% core contribution at 5\,GHz. The radio images \citep{Spencer1991,Ludke1998} indicate multiple features that require a double power law in our fitting procedure (see Table \ref{tab:best_radio_fit}). As shown in Figure \ref{fig:sed}, the radio emission cutoff occurs at FIR wavelengths. While this suggests that the non-thermal emission may contribute significantly at submm wavelengths (hence SFR estimates based on submm data should be treated carefully), submm observations are required for constraining the radio fit.

The emission from the torus is best described with a combination of clumps and a homogeneous disk with maximum acceptable opacity in the torus library (see Table \ref{tab:sieb_par}). This, combined with a small filling factor and 43$^{\circ}$ inclination angle, results in moderate obscuration of the visible-UV emission of the accretion disk. While quasar spectra, in general, show silicate emission features, the rest-frame 9--16 \um spectra of 3C 190 indicate the presence of silicate absorption \citep{Leipski2010}. In our fitting procedure, we examined the torus templates with silicate absorption; however, none of those results in a better fit than that shown in Figure \ref{fig:sed}. 

Comparing the best-fit accretion disk with the X-ray data does not indicate the presence of an underlying X-ray component triggered by radio structures.


\textbf{3C 191} was classified as a CSS in some earlier studies \citep{Akujor1995,Willott2002},  however, using the classification of \pece\ and radio jets of 81 kpc we do not classify this object as a CSS. 3C 191 has $\sim$9\% core contribution at the 5 GHz, and its radio emission is best described as a single power-law with a cutoff at FIR wavelengths (see Table \ref{tab:best_radio_fit}). Therefore, as illustrated in Figure \ref{fig:sed} the submm data and likely FIR  have a significant non-thermal contribution. 

The emission from the torus is best described with a combination of clumps with a low volume filling factor and a homogeneous disk with moderate opacity and an inclination angle of $43^{\circ}$, which implies no significant obscuration of the accretion disk emission. The low level of obscuration is consistent with the optical spectral slope, $\rm \alpha_{opt}$, of 0.7 \citep{Barthel1990} that is a typical value of radio quasars with little obscuration \citep[see also ][]{Brotherton2001,Willott2002}.

3C 191 spectra indicate the presence of a strong absorbing system (rest-frame equivalent width of 6.1 \AA) associated with {\ion{C}{4}} absorption line \citep{Anderson1987} $\sim$ 30 kpc from the nucleus \citep{Hamann2001}. Studies find that quasars with strong associated absorption lines are in the early stage of their development, and smaller/younger radio structures are more common to have associated absorption lines \citep[e.g.,][]{Becker2000,Becker2001,Willott2002}. \maz{However, as discussed in detail in \cite{Barthel2017}, the absorption feature in 3C~191 is possibly connected to starburst-driven superwinds \citep[see also][]{Hamann2001}.}

In most of our sources (see Figure \ref{fig:sed}), the torus component dominates the blue side of the FIR bands; interestingly, in 3C 191, it dominates over the host galaxy emission at all the MIR-FIR bands. The Herschel upper limits in the FIR allow for the possibility of a weak, FIR, cool dust contribution below the submm waveband where the synchrotron component dominates. 
\reff{3C 191 was classified as a hyperluminous quasar with $\rm L_{FIR} > 1.2 \times 10^{13} L_{\odot}$ \citep{Willott2002}. However, we find the torus as the dominant source of IR emission rather than the host galaxy.} 

Finally, we note that the best-fit accretion disk model indicates the presence of a non-thermal X-ray component possibly associated  by radio structures.

%

\textbf{3C 204} has an extended jet of $\sim$543 kpc and $\sim$ 8\% core contribution in the 5 GHz band. The quasar radio SED is well fitted with a broken power-law (see Table\ref{tab:best_radio_fit}). The VLA 4.9 GHz images show multiple features, including a bright radio core, hot spots, and a one-sided jet that deflects towards the end \citep{Bridle1994}. There is some evidence of [{\ion{O}{2}}] emission perpendicular to the jet axis \citep{Bremer1992,Bridle1994}
at $\sim$ 90 kpc towards the north and  $\sim$ 45 kpc towards the south.

The emission from the torus is best described with a combination of clouds with negligible optical depth and a homogeneous disk with moderate opacity. This, combined with an inclination angle of $52^{\circ}$ results in no significant obscuration of accretion disk at visible-UV.

Comparing the best-fit accretion disk model with the X-ray data suggests the presence of additional, non-thermal X-ray emission possibly triggered by radio jets.


\textbf{3C 205} has extended radio structures of $\sim 251$ kpc with a core contribution of 3\% in the 5GHz band. The radio emission is best described as a broken power-law (see Table \ref{tab:best_radio_fit}), which is consistent with the presence of the multiple radio structures \citep{Lonsdale1984,Lonsdale1986}.

The emission from the torus is best described with a combination of clumps with small opacity and a homogeneous disk with moderate opacity, which result in a small amount of obscuration of the accretion disk radiation. 
3C 205 spectra indicate the presence of a strong absorbing system (rest-frame equivalent width of jhc3.21 \AA) associated with {\ion{C}{4}} absorption line \citep{Anderson1987}.

The best-fit accretion disk model predicts the presence of little non-thermal radiation at the X-ray band triggered by radio structures.


\textbf{3C 208} has a radio jet of $\sim 172$ kpc with $\sim$ 10\% core contribution at 5 GHz. The radio emission is best fitted with a broken power-law that is consistent with the 
presence of multiple radio features, including a 
bright core and the hot spots \citep{Bridle1994}.
3C 208 has a one-sided jet that is straight for most of its length but deflects towards the end \citep{Bridle1994}.

The emission from the torus is best described with a combination of clumps with a small optical depth  and a homogeneous disk with moderate opacity, which result in no significant obscuration at visible-UV wavelengths.
The available \textit{HST} images (F606W and F140W bands) do not indicate any evidence of nearby merging sources \citep{Hilbert2016}.

The X-ray spectrum fitted by the accretion disk model implies some contribution from the radio structures.


\textbf{3C 212} is a red quasar \citep{Aldcroft2003} with radio jets of $\sim$ 216 kpc and $\sim$ 17\% core contribution at 5 GHz.
MERLIN observations at 6 cm and 15 cm  \citep{Akujor1991} indicate multiple features including a bright core that best fitted with the fourth model in our fitting procedure (see Section \ref{sec:radio} and Table \ref{tab:best_radio_fit}). 
\reff{In this case, the cyan component indicates the emission from the extended radio structures/hot spots, and the dark blue component illustrates the emission from the core. We note that each component is truncated at its own cutoff frequency.}
To constrain the radio fit we used recent ALMA observations (see Table \ref{tab:submmdata}), which are dominated by the non-thermal radiation from the radio structure (Figure \ref{fig:sed}). 

Consistent with its classification as a red quasar, our torus model indicates a combination of clumps and a homogeneous disk with moderate optical depth. These components result in moderate obscuration of the accretion disk emission. 
This target also has X-ray and UV absorbers \citep{Aldcroft2003}.

The X-ray spectrum fitted by the accretion disk model implies the presence of an X-ray component possibly triggered by the radio structures.


\textbf{3C 245} is a moderately beamed quasar \citep[with a jet oriented at $<20^{\circ}$ to the line of sight, ][]{Foley1990,Marin2016} with $\sim 1067$ kpc jets and $\sim 66\%$  core contribution at 5 GHz, which is significantly higher than other sources in our sample (see Table \ref{tab:radio}). \reff{The radio emission is best described as a parabola (shown in cyan) with an additional core component (shown in dark blue)}. 
To constrain the radio contribution to the FIR, we recently obtained the SMA data (see Table \ref{tab:submmdata}), which is dominated by the non-thermal radiation from the radio structure (Figure \ref{fig:sed}). \reff{The extreme variability expected in blazars may not be seen in 3C 245 since our recent SMA observation is consistent with older radio data \citep{1981AJ.....86.1306G}.}

Consistent with its classification, the best-fit torus
indicates a combination of clumps with negligible opacity and a homogeneous disk with small opacity. This combination results in no significant obscuration from the clouds and the homogeneous dusty disk. The torus inclination angle  ($19^{\circ}$, see Table \ref{tab:best_fit_torus}) and the accretion disk inclination angle ($26^{\circ}$) are consistent with the radio-determined inclination angle \citep{Foley1990,Marin2016}.

The best-fit accretion disk model also indicates the presence of significant non-thermal X-ray emission possibly triggered by the radio structures. We note that since 3C 245 has nearly a face-on inclination angle and significant underlying non-thermal radiation at radio/mm and X-ray wavelengths, the non-thermal continuum may contribute significantly to the IR-visible-UV bands as well. 
Given the significant non-thermal emission and quasar's dominance at visible-UV bands, it is unlikely to have a reliable estimate of the host galaxy properties from our SED fits.


\textbf{3C 268.4} has jets of 205 kpc
and $\sim8$\% core contribution at 5 GHz. The radio images \citep{Lonsdale1986,Liu1992}
indicate multiple features, including a bright core and double hot spots. The radio emission is best fitted with a double power-law model with a cutoff at FIR wavelengths (see Table \ref{tab:best_radio_fit}), resulting in significant non-thermal contribution at submm wavelengths \citep{Willott2002}.

The emission from the torus is best described with a combination of clumps and a homogeneous disk with the highest acceptable opacity in the torus library (see Table~\ref{tab:sieb_par}). However, this, combined with a small filling factor and inclination angle of $52^{\circ}$, results in little obscuration 
of the accretion disk emission. Similar to 3C 191, 3C 268.4 is also classified as a hyperluminous quasar with $\rm L_{FIR} > 2 \times 10^{13} L_{\odot}$ \citep{Willott2002}. \reff{Similar to 3C~191, we find more contribution from the torus than the host galaxy in the 160 \um and 350 \um bands.} We also note that similar to 3C~191, the spectra of 3C~268.4 show strong {\ion{C}{4}} associated absorption  \citep{Anderson1987}.

3C 268.4 was reported as a lensed quasar with a foreground cluster at $z\sim0.35$ \citep{Sanitt1976}. \maz{We did not correct the SED for the magnification by the foreground cluster.} Recent \textit{HST} observations indicate the presence of a bright star-forming clump $2.5"$ from the center and an additional oblong source with both optical and radio emission $0.8"$ from the center \citep[see Figure 10 in ][]{Hilbert2016}. Since the IR emission from these structures cannot be resolved from AGN emission with the current data, the hyperluminous quasar classification should be treated with caution (see 3C 318 below). 

The X-ray spectrum fitted by the accretion disk model implies no significant contribution from the radio structures.

\textbf{3C 270.1} has a jet of $\sim$374 kpc and a relatively high core contribution (22\%) at 5 GHz. Multi-frequency radio images of 3C 270.1 indicate the presence of a strong core and hot spots \citep{Liu1992} and a one-sided jet \citep{Hilbert2016}. The radio emission is best fitted with a double power-law with a cutoff at FIR wavelengths resulting in significant non-thermal contribution at mm wavelengths.

The emission from the torus is best described with a combination of clumps with a small filling factor and a homogeneous disk with the highest acceptable opacity in the torus library. This, combined with a $52^{\circ}$ inclination angle, results in little reddening of the accretion disk emission. 

3C 270.1 spectra indicate the presence of a strong absorption complex (rest-frame equivalent width $>6.17$ \AA) associated with {\ion{C}{4}} absorption line \citep{Anderson1987}. 

The best-fit accretion disk model indicates the presence of significant  emission at the X-ray waveband possibly triggered by radio structures. \textit{Chandra} observations find extended X-ray emission that is co-spatial with the radio lobe and peaks at the position of the hot spots \citep{Wilkes2011}. The extended X-ray emission is not included in the X-ray data used for this SED analysis.


\textbf{3C 287} is a CSS with projected radio jets of $\sim$ 8 kpc. Because the jet inclination angle for this object is unknown, we were unable to estimate its de-projected jet length. Also, we were unable to find the $\rm R_{CD}$ value for this object, therefore adopted the average value of the CSS quasars in our sample (Section \ref{sec:radio}), which resulted in $\sim$10\% core contribution. The radio emission is modeled as a double power-law with a cutoff at submm/FIR wavelengths (see Table \ref{tab:best_radio_fit}), resulting in significant non-thermal contribution at mm-submm wavelengths. The VLBI and MERLIN observations indicate the presence of multiple radio structures, including a curving jet \citep{Fanti1989}.

The emission from the torus is best described with a combination of clumps with a negligible optical depth and a homogeneous disk with the highest acceptable opacity in the torus library (see Table \ref{tab:sieb_par}). This combination with the inclination angle of $19^{\circ}$ ($26^{\circ}$ from the accretion disk model) result in no significant obscuration of the visible-UV radiation from the accretion disk. 
While the extreme variability is expected in moderately beamed quasars, it may not be seen in CSS and GPS sources such as 3C 287 since they are young \citep[e.g.,][]{Salvesen2009}.

The X-ray spectrum fitted by the accretion disk model implies a significant contribution associated with the radio structures. The X-ray data obtained with XMM-Newton and \textit{Chandra} \citep{Salvesen2009,Wilkes2013} find a soft X-ray spectrum ($\Gamma$=1.8) that can be fitted with a simple power-law. 

Similar to 3C 245, 3C 278 is viewed almost face-on and has significant underlying non-thermal radiation at radio/submm and X-ray wavelengths. The non-thermal continuum may contribute to other wavelengths. 
We also note that the \textit{HST} images indicate a few nearby sources \citep[$\sim 5"$,][]{Hilbert2016}, and the IR emission may be contaminated by one/more of these sources. Altogether, these make the host galaxy properties estimated from our SED fits uncertain.

\textbf{3C 318} is a CSS with jets of $\sim 8$kpc and core contribution $<12$\% at 5 GHz. The 18 cm MERLIN and  VLBI observations \citep{Spencer1991} show a two-sided jet, which fades before reaching the lobes. The emission from the radio structure is best described with a double power-law with a cutoff at 200 GHz. In addition to the 1.2 mm MAMBO data, we used recent ALMA observations 
\citep[see Table \ref{tab:submmdata} and ][]{Barthel2019} to constrain the radio model. \cite{Barthel2019} used the 2 cm VLA image to subtract the non-thermal emission from the ALMA image at 1mm  and estimated that $\sim$11\% of the total flux at 1 mm has a non-thermal origin. 
We use this estimate to constrain the radio model in our fitting procedure. However, with this prior estimate, the cutoff in our fitting procedure happens at $\sim 200$ GHz; consequently, we estimate no non-thermal contribution at 1 mm. Therefore there could be $\lesssim 10\%$ non-thermal contribution at submm wavelengths \reff{\citep[also see][]{Haas2006}.}

The torus emission is best described with a combination of clumps and a homogeneous disk, both with high opacity. However, due to the small filling factor of the clumps  and the 33$^{\circ}$ inclination angle, these components do not result in a significant obscuration of the accretion disk emission. 

The best-fit accretion disk indicates the presence of an additional X-ray component possibly triggered by radio structures. We note that with only a few data points at optical, UV, and X-ray range, the physical parameters driven from the accretion disk model (see Table \ref{tab:best_fit_torus}) should be treated with caution.

\reff{We note that 3C 318 is classified as a hyperluminous infrared quasar with $\rm L_{FIR} > 10^{13} L_{\odot}$ in some earlier studies \citep{Willott2002,Willott2007}. Recently, \cite{pece2016} reported that most of the fluxes measured in earlier studies originate in a pair of bright interacting galaxies at $z\sim0.35$. To robustly estimate the AGN and host galaxy properties, we used the fluxes from Table 1 in \cite{pece2016}. After subtracting the contamination of the nearby source from the photometry, 3C 318 has an SFR of $\sim 320 ~M_{\odot}/$yr, a factor of 5 lower than the \cite{Willott2007} estimation.}

\textbf{3C 325} was originally classified as a radio galaxy and later was reclassified as a red quasar \citep{Grimes2005} based on optical spectroscopic data. 
This quasar has a projected radio jet of 124 kpc and less than 1\% core contribution at 5 GHz. Because the jet inclination angle for this object is unknown, we were unable to estimate its de-projected jet length. The VLA images show multiple features, including bright hot spots and asymmetrically placed lobes \citep{Fernini1997}. 
The radio data are best fitted with a double power-law, which turns down before reaching the submm wavelengths.

Consistent with its X-ray spectral analysis \citep[$ \rm N_H\sim6.2\times 10^{22}cm^{-2}$,][]{Wilkes2013}, the emission from the torus is best described with a combination of 
clumps and a homogeneous disk with moderate opacity. This combination results in moderate obscuration of the accretion disk emission. Considering the lack of data in 1--3 \um (observed-frame), the best-fit accretion disk is determined with the data points at shorter wavelengths; therefore, the accretion disk parameters are not well constrained.

The best-fit accretion disk does not imply the presence of non-thermal X-ray emission. The X-ray analysis indicates a moderately hard X-ray spectrum with the hardness ratio (H-S/H+S, where H and S are the net count rates in the 2--8~keV and 0.5--2~keV X-ray bands) of +0.05,  which is harder than all the other quasars in our sample \citep{Wilkes2013}.

\textbf{4C 16.49} extends over 216 kpc with $\sim$ 5\% core contribution at 5 GHz. The radio images obtained with VLA at 2 and 6 cm 
show a strong radio core, jet, and a small counter-jet \citep{Lonsdale1993}. The radio emission is best described with a double power-law (see Table \ref{tab:best_radio_fit}). \reff{Our model does not predict any non-thermal contribution from the radio structures at shorter wavelengths, however, due to the small number of reliable data points and lack of mm and submm data, we should be cautious in the interpretation of the results}. 

The best-fit torus model combines clumps with  negligible opacity and a homogeneous disk  with  the highest acceptable opacity in the torus library (see Table \ref{tab:sieb_par}).  This combination results in no significant obscuration of visible-UV emission from the accretion disk.  

The best-fit accretion disk template also confirms the presence of a significant X-ray component possibly triggered by radio structures.
However, unlike other sources in our sample, 4C 16.49 do not have SDSS or other recent reliable data at visible-UV  wavelengths and is only limited to old SuperCOSMOS observations. 
Given the limited number of visible-UV photometric measurements, the accretion disk may not be well-constrained, and the X-ray component from the radio structures may not be significant.


\textbf{3C 432} 
extends over $\sim155$ kpc and has a core contribution of $\sim3$\% at 5 GHz. The VLA images indicate multiple structures, including radio lobes, bright hot spots, and a one-sided jet \citep{Bridle1994}. The emission from these radio structures is best described with a double power-law (Table \ref{tab:best_radio_fit}).

The torus emission is best fitted with a combination of clumps with negligible opacity, and a homogeneous disk with high opacity that are viewed at $33^{\circ}$ inclination angle
and result in no significant obscuration of the accretion disk emission. The \textit{HST} F606W and F140W images show extended narrow line regions (within $8"$) along the direction of the radio lobes \citep{Hilbert2016} with several faint sources within this radius, which could potentially contaminate the quasar SED.

The X-ray spectrum fitted by the accretion disk implies no contribution from the radio structures to the X-ray emission.


\textbf{3C 454.0} is a CSS with de-projected jets of $< 46$ kpc and $<25$\% core contribution at 5 GHz. The radio images show various features, including the core and hot spots \citep{Spencer1991,Ludke1998}.
The radio emission is best described with a parabola having a cutoff at FIR wavelengths. To constrain the radio model we added recent ALMA data to our analysis \citep[see Table \ref{tab:submmdata} and ][]{Barthel2019}. We used the non-thermal emission estimated in \cite{Barthel2019} as prior information to constrain the radio model in our fitting procedure. Similar to \cite{Barthel2019} we estimate $\sim 82$\% non-thermal emission at 1mm.

The torus emission is best fitted with a combination of clumps with small opacity and a homogeneous disk with the maximum acceptable $\rm \tau_{D}$ in the torus library (see Table \ref{tab:sieb_par}). This combination at an inclination angle of $43^{\circ}$ results in no significant obscuration of the accretion disk emission. 

The X-ray spectrum fitted by the accretion disk implies contribution from the radio structures at X-ray wavelengths.

\subsection{Commonalities Among Sources} \label{sec:common}
\label{sec:fit_summary}
\begin{itemize}
\item Out of the 20 quasars in our sample, 11 (3C 43/181/190/191/212/245/268.4/270.1/287/454.0) have $>90\%$ non-thermal contamination at 1.25\,mm (see Section \ref{sec:non_th} for more details). \maz{While the mm--submm emission in some of these sources is dominated by the core (3C~212/245), in others it can be dominated by the  extended structures and hot spots.}

\item 
\reff{In 13 sources (3C9/43/191/204/208/212/245
/268.4/270.1/287/318/454.0 and 4C 16.49) there are discrepancies between the best-fit accretion disk and the X-ray data. Different factors may contribute to these discrepancies. X-ray emission associated with radio structures may be present, especially for sources with high $\rm R_{CD}$ (e.g., 3C 245/270.1).  This may confirm a jet-related contribution to the X-ray emission in radio-loud quasars and may contradict findings \citep[see ][]{Zhu2020} which attribute the X-ray brightness of the radio-loud quasar (relative to the radio-quiet population) to the corona rather than the jet. Potential contamination from the extended X-ray emission in our X-ray luminosity estimation and underestimation of the X-ray emission in radio-loud AGN in the \cite{Kubota2018} accretion disk model are also possible factors contributing to the discrepancies.}

\item In sources with small inclination angles (3C 245/287), the significant underlying non-thermal radiation at radio--submm and X-ray wavelengths may contribute at
IR, visible, and UV bands as well. Therefore, the torus and the host galaxy properties derived from SED fitting are uncertain.

\item The \textit{HST} \citep[F606W and F140W bands, ][]{Hilbert2016} of some of our sources (3C 268.4/287 and maybe 3C 432) 
indicate the presence of a few nearby objects, which may contaminate the photometry. \reff{In these cases, the physical properties derived from SED fitting are poorly unconstrained.}

\item Four quasars in our sample 3C 191/205/268.4/270.1, with de-projected radio jets of 81, 251, 205, and 374 kpc have associated \ion{C}{4} absorption complexes \citep{Anderson1987} with rest-frame equivalent width of 6.12 \AA\ , 3.21 \AA, $>1.87$ \AA\ and $>6.17$ \AA, respectively. While some studies \citep[e.g.,][]{Becker2000,Becker2001,Willott2002} suggest that associated absorption is more common in sources with smaller/younger radio structures, there does not appear to be such a relationship in our sample.

\item Our SED modeling is limited by the available photometry (see Section \ref{sec:limit}). In sources with few reliable visible-UV data points (3C 14/318/325 and 4C 16.49), the accretion disk parameters may be poorly  constrained.

\item \maz{Extinction in the red quasars  (3C~14/190/212/325) is primarily due to a highly inclined torus rather than the dust extinction in the host galaxy.}

\end{itemize}
\section{Discussion} \label{sec:discussion}

\begin{figure*}[t] 
    \includegraphics[width=0.42\textwidth,angle =90]{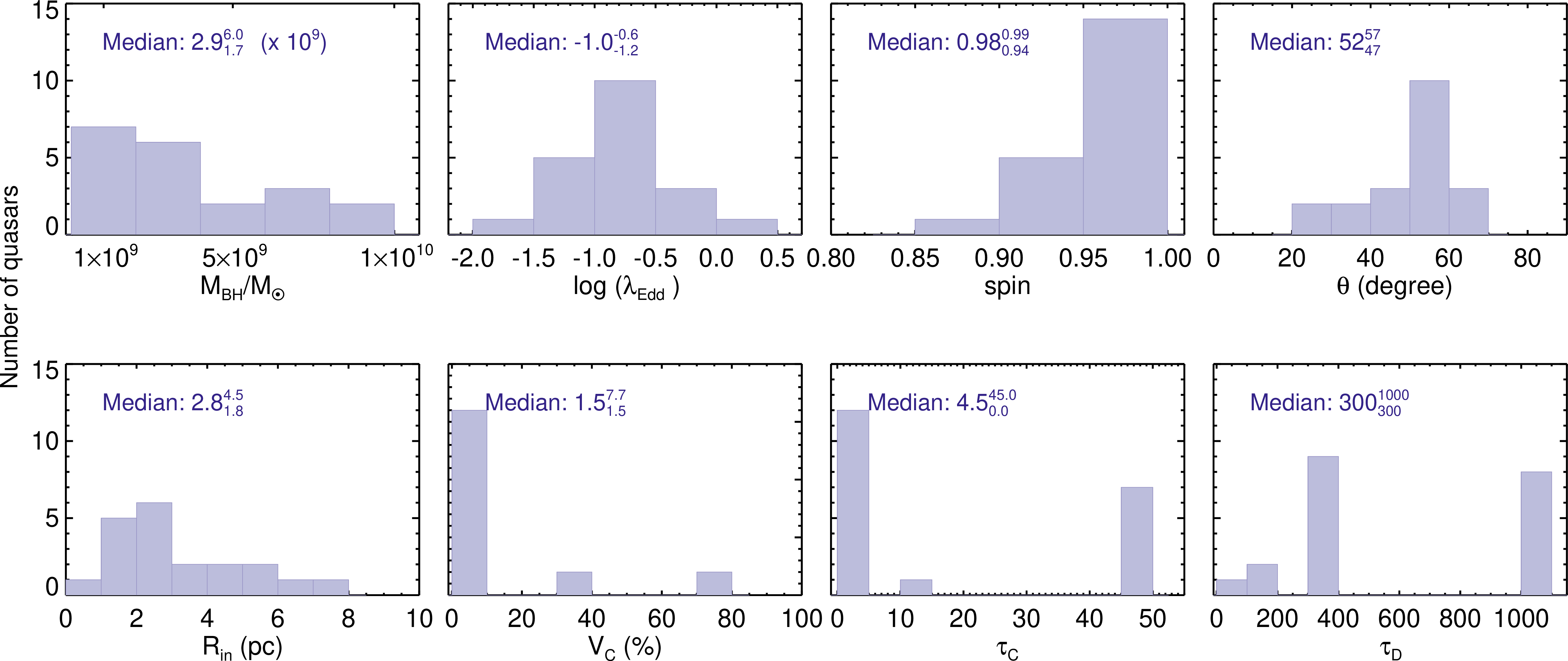}
             \caption{The distribution of the physical properties of the AGN derived from the
             accretion disk and the torus fits for our 3CRR quasars, along with the median values in each case and their associated 25th (subscript)-75th (superscript) percentile ranges.}
\label{fig:agnhist}
\end{figure*} 

In this study, we present a state-of-the-art AGN radio-to-X-ray SED model (ARXSED) that decomposes radio-loud quasars into their AGN and host galaxy components by fitting their photometry over ten decades in frequency space.  With ARXSED, we investigate the properties of a sample of 20 3CRR quasars at \z.  Below we first discuss the uniqueness and limitations of our technique, as well as the average properties of the AGN in our sample.  We then present the median SED of the radio-loud quasars obtained with our fitting technique. Finally, we compare the details of our model, the median SED, and the properties of our quasars with the literature.

\subsection{Uniqueness, Limitations, and Biases of ARXSED} \label{sec:limit}

An advantage of this study compared to previous work \citep[e.g.,][]{JK2003, Mullaney2011, pece15,pece2016} is the wavelength coverage. ARXSED treats photometry extending from radio to X-ray wavelengths. Specifically, ARXSED models the radio emission (Section \ref{sec:radio}), the torus (Section \ref{sec:torus}), and the accretion disk (Section \ref{sec:AD}) as well as the host galaxy (Section \ref{sec:magphys}). Thus it simultaneously accounts for radiation from different structures surrounding the SMBHs. While ARXSED considers a component that accounts for the host galaxy emission from radio to UV wavelengths, some studies rely on scaling relations such as $\textrm M_{\rm BH}-\sigma$ 
\citep[e.g.,][]{marconi2003relation}, $\textrm L_{\rm host}-L_{\rm AGN}$ \citep{Vanden2006} or color-color diagnostics \citep[e.g.,][]{Grewing1968, Sandage1971, Elvis2012} to account for the host contribution. Considering the uncertainty and large scatter in each of these scaling relations, the physical properties of the AGN derived from their analysis may not be robust.

Like most modern SED fitting codes, ARXSED implements a self-consistent approach to dust attenuation, in which the \emph{intrinsic} SED of radio-loud quasars is appropriately corrected for the reddening and absorption occurring in the torus, the host galaxy, and along the line of sight in the Milky Way.

The main limitations of our approach are a lack of reliable photometry.
Several sources in our sample (e.g., 3C\,325, 4C\,16.49) lack reliable visible-UV photometry and/or have very few data points, leaving their  accretion disk parameters poorly constrained.

ARXSED uses the thermal visible-to-X-ray continuum fitting \citep{Zhang1997} technique to constrain the SMBH and the accretion disk properties.
Constraining the accretion disk parameters without observations around the peak of the thermal continuum is challenging. The peak of the accretion disk is sensitive to the BH mass, Eddington ratio, and spin (see Figure \ref{fig:QSOSED}), therefore, not having good constraints results in larger uncertainties. To accurately determine the peak of this thermal continuum (occurring around 100-1000\AA\ rest-frame for SMBHs), we require far-UV (FUV) observations from 
space (e.g., the Cosmic Origins Spectrograph on \textit{HST}), and soft X-ray data contributing to the high energy part of the thermal continuum.


Another advantage of ARXSED is that it implements multi-component radio models that account for a steepening or cut-off due to the aging of the electron populations. We find that a single power-law ($L_{\nu} \propto \nu^{\alpha}$) can not adequately model the radio emission when compact structures like
cores and hot spots are present. 
In addition, long-wavelength radio photometry usefully constrains the non-thermal radiation from the radio structures at shorter wavelengths.  However, a lack of submm data and high S/N \textit{Herschel} observations may result in an underestimation of the non-thermal contamination in some cases (e.g., 3C\,204) and a corresponding overestimate of SFR.

\reff{Given a large number of parameters and diverse data quality, we apply self-consistency checks (e.g., tie the inclination angle of the torus and accretion disk) and use published information (e.g., $M_{\rm BH}$ estimates from emission lines) as priors to the SED fits. However, similar to any SED fitting code, the best-fit SEDs are not unique, and derived parameters may be degenerate with one another. We also note that AGN variability, in particular at radio and visible-UV wavelengths, can bias our SED fits,
and impact the physical parameters constrained from the model. To minimize the impact of variability at different wavelengths, we only kept the observations that were close to each other 1) in flux and 2) in observing time.}
 
ARXSED is based on the assumption that
the photometry is dominated by the radiation from the AGN at most wavelengths (except for the FIR; see Section \ref{sec:fitting}); therefore, we may underestimate the stellar mass in some sources. 
Also, in sources with small inclination angles (3C\,245 and 3C\,287, see Table \ref{tab:best_fit_torus}) beamed non-thermal emission at IR-visible-UV wavelengths, unaccounted in our modeling, add uncertainty to the derived host galaxy properties. We will discuss the results and the limitations of our technique for SFR and stellar mass measurements in detail in a following paper (Azadi et al., in prep).

\subsection{Physical Properties of the SMBHs \\
and Dusty Tori in 3CRR Quasars} \label{sec:3CRsed}

The distributions of the SMBH, torus and accretion disk properties derived from the SED fits for our 3CRR quasars are shown in Figure.~\ref{fig:agnhist}, along with the median values in each case. The BH mass, Eddington ratio, and spin are constrained by the best-fit accretion disk (see Table \ref{tab:best_fit_torus}). The inner radius of the torus, the volume filling factor, the optical depths of the dust clouds, and the homogeneous dusty disk are constrained by the torus model. The inclination angle (measured from the pole) is the average value from the best-fit torus and the accretion disk.

The average BH mass (and the standard deviation) of the quasars in our sample is $(3.7\pm 2.7)\times10^{9} M_{\odot}$. \reff{Our BH masses are obtained from the best-fit accretion disk templates, while the templates are built based on prior mass estimates from broad  {\ion{C}{4}~$\lambda$1548} or {\ion{Mg}{2}~$\lambda$2800} emission lines (Section \ref{sec:AD}) allowing the mass to vary within $\pm$ 1 dex of the initial estimates (see Table \ref{tab:qsosed}).}
Using {\ion{C}{4}~$\lambda$1548}, {\ion{Mg}{2}~$\lambda$2800} or \hb, \cite{McLure2006} estimated $ M_{BH}$ for 18 out of 20 quasars in our sample (except for 3C 318 and 3C 325) and obtained an average of 
$(3.8\pm 2.9)\times10^{9}  M_{\odot}$. Indeed our estimates of the 3CRR BH masses are consistent with the BH mass estimates of the non-3C radio-loud AGN at similar redshifts \citep[e.g.][]{Liu2006} but slightly higher than those at lower redshifts \citep[e.g.][]{Coziol2017}.
\reff{Recently, \cite{Collinson2015, Collinson2017} investigated the intrinsic NIR to X-ray SED of 11 radio-quiet quasars at $1.6<z<2.2$ and using the \ha\ line estimated an average BH mass of $(1.5\pm 1.4)\times10^{9} M_{\odot}$ which overlaps with the mass of the BHs in our sample.
Consistently, with a larger sample at $z<0.5$, \cite{McLure2002} found that radio-loud quasars have larger BH masses than their radio-quiet counterparts, although with a large overlap.}


The average quasar $\log(\lambda_{Edd})$ obtained for our sample is $-0.87\pm$0.41 using the QSOSED templates of \kd. However, lack of the FUV and soft X-ray data can bias the measurements toward more massive black holes and/or lower Eddington ratios (Section \ref{sec:AD}). Adopting the SMBHs masses from \cite{McLure2006}, and the bolometric correction from 
\cite{Heckman2004} for estimating 
$L_{bol}$ from $L_{[\rm OIII]}$, recently \cite{Daly2019} estimated the average $\log \lambda_{Edd}$ for 15 of the quasars in our sample to be $-0.32\pm$0.42. Given the scatter in the average values, the lower $\log(\lambda_{Edd})$ in our sample may be attributed to a different methodology used in estimating the Eddington ratio. 
\reff{Adopting the OPTXAGNF accretion disk model of \cite{Done2012},  \cite{Collinson2015} estimate an average of $\log(\lambda_{Edd})$ of $0.02\pm0.57$ for radio-quiet quasars at similar redshift. 
\cite{Coziol2017} with a large sample of radio-quiet and loud sources at $z<0.3$ found comparable Eddington ratios for the two populations.
Overall, we find a lower Eddington ratio than radio-quiet sources at similar redshifts, with some overlap. But the fact that our best-fit accretion disk model is mostly driven  by visible-UV photometry may result in an underestimation of the Eddington ratio (by  moving the peak of accretion disk SED to lower frequencies, see Figure \ref{fig:QSOSED}).}

Another parameter estimated from our accretion disk modeling (i.e., thermal continuum fitting) is the SMBH spin. Although the continuum fitting method can estimate the spin of stellar-mass BHs, its application to AGN can be more challenging. In AGN, the inner accretion disks are relatively cool ($T \sim 10^5$ K), with the bulk of the thermal emission occurring in the far-or-extreme UV regime, which is hard to observe from the ground. However, recent studies of SMBHs with well-constrained masses found that spins estimated with the thermal continuum fitting method are in good agreement with estimates from reliable techniques such as the X-ray reflection technique \citep[e.g.,][] {Capellupo2016}. 
The average spin of the quasars estimated from the best-fit accretion disk is $0.97\pm0.04$. Adopting the OPTXAGNF accretion disk model \citep{Done2012} \cite{Collinson2017} found their radio-quiet sample quasars have spins $<0.9$. While radio-quiet quasars are expected to have lower spin compared to radio-loud sources, they still may have high spin values (see \citealt{Brenneman2006}). Adopting \cite{BZ1977} framework in which the BH spin is related to the magnetic field, \cite{Daly2019} estimated the spin of 750 SMBHs and, for the 15 quasars included in our sample, found an average spin of 0.99$\pm$0.01 \citep[see also ][]{Daly2011}.

\begin{figure*}[t]
\begin{center}
    \includegraphics[width=0.32\textwidth ,angle =90]{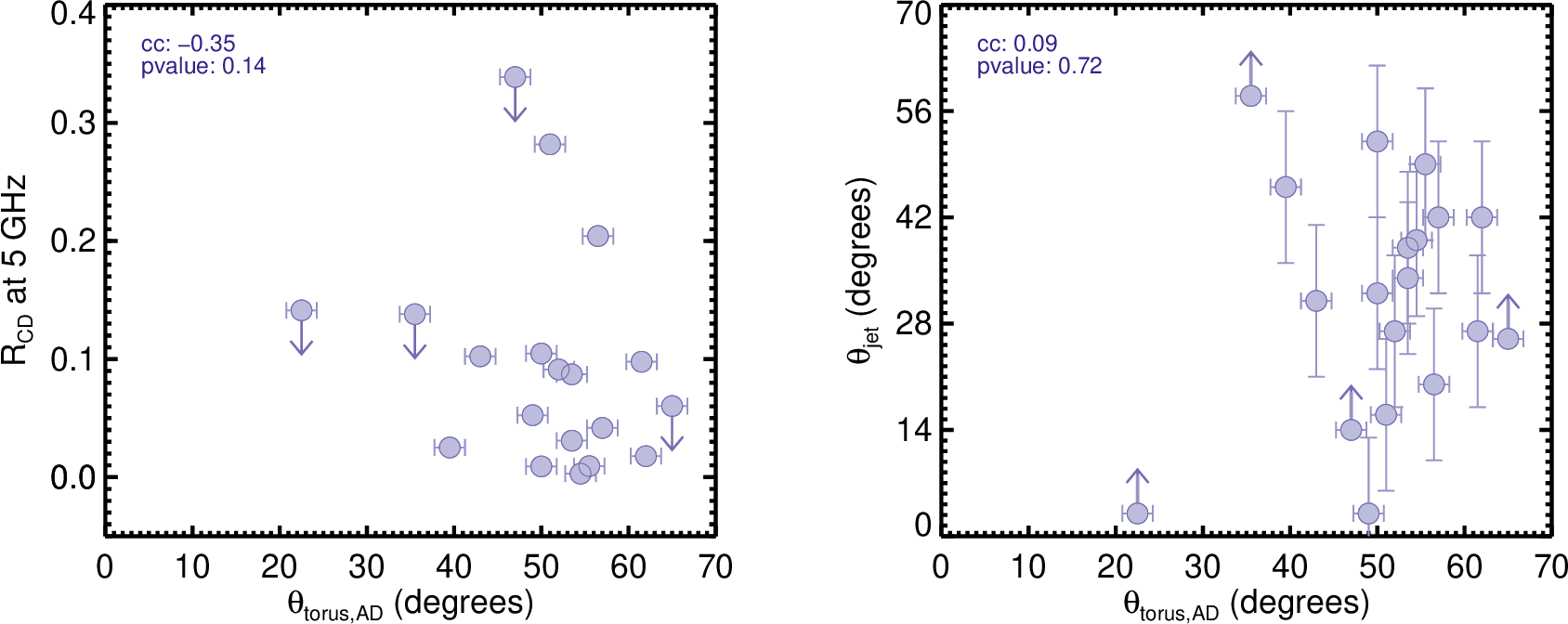}
             \caption{\reff{\textit{Left:} The relation between the radio core dominance $\rm R_{CD}$ and the average inclination angle of the torus and accretion disk. \textit{Right:} The relation between the jet inclination angle and the average inclination angle of the torus and accretion disk. The correlation coefficient and its significance level are reported in each panel. For a better illustration, we removed 3C~245 which is an outlier with $\rm R_{CD}$ of $\sim 1.95$.}}
             \label{fig:rcd}
\end{center}

\end{figure*} 

\reff{The best-fit torus parameters indicate low obscuration, consistent with Type 1 nature of our sources.} Indeed all the sources with a high filling factor in Table \ref{tab:best_fit_torus} have clumps with negligible optical depth.  
Also, the negligible optical depth of the clumps in 7 out of 20 quasars in our sample (Table \ref{tab:best_fit_torus}) indicates that the primary source of torus obscuration is dust in the toroidal disk rather than the clumps. For red quasars (3C\,14/190/212/325) the obscuration of the visible-UV emission from the accretion disk is more pronounced (Figure \ref{fig:sed}), as expected.

The average inclination angle (measured from the pole) obtained from the best fit torus and accretion disk is 49$^{\circ} \pm 12^{\circ}$, which is consistent with other studies of non-blazar radio-loud (broad-lined) AGN \citep[e.g.,][]{Willott2000,Arshakian2005,Marin2016}.
\reff{\cite{Barthel1989} found an average inclination angle of 31$^{\circ}$ for 3CRR quasars at $z<1$ (estimated from $R_{\rm CD}$), and 45$^{\circ}$ as the division between quasars and narrow-line radio galaxies, using the jet inclination angle. We present the inclination angle from the best fit SED (average of the torus and accretion disk) in Figure \ref{fig:agnhist}, while the average viewing angle from the $R_{\rm CD}$ in our sample is $\sim 33^{\circ}$ (including the limits). We note that our sample i) does not include blazars\footnote{\maz{Our sample is complete in radio orientation which means that if there was a 3CRR blazar within this redshift range it would be included. However, none of our sources have inclination angle $< 20^\circ$.
}}, ii) includes sources with potential non-thermal contamination from the radio structures at IR (e.g., 3C~43), so that the shape of their IR SED (and consequently the best-fitted torus model) is affected by that, iii) includes red quasars for which, to replicate their IR SED, we need torus templates with higher dust content and/or larger inclination angle. Indeed all the sources with torus inclination angle $>60^{\circ}$ in our sample are either red quasars (3C~14/190/212/325) or potentially have more non-thermal emission at IR wavelengths than what our model predicts (3C~43/212). We discuss the discrepancies between radio and accretion disk/torus measurements below in Section \ref{sec:orientation}.}

\subsection{The Orientation of 3CRR Quasars} \label{sec:orientation}

\reff{In Figure \ref{fig:rcd} we investigate the relation of the radio core dominance, $\rm R_{CD}$ (Section \ref{sec:radio}),
at 5\,GHz and the average inclination angle of the torus and accretion disk obtained from our best fit, as well as the relation between the jet and torus/accretion disk inclination angles. In each panel we report the correlation coefficient and its significance level based on \texttt{r-correlate} \footnote{ \texttt{r-correlate} computes the Spearman’s rank correlation coefficient (cc) and the significance of its deviation from zero (pvalue). A correlation is considered significant if pvalue $<0.05$. In this case, it is unlikely for the correlation to have occurred by accident.} routine in IDL. For better illustration, we removed 3C~245, which is an outlier in our sample 
with a core contribution of 66\% and $\rm R_{CD}$ of 1.95.}

\reff{We do not find any statistically significant relation between the two parameters in either of the two panels of Figure~\ref{fig:rcd} after removing 3C~245 (with 3C~245, we find cc = -0.44, pvalue = 0.05 in the left panel). 
The right panel of Figure \ref{fig:rcd} shows the relationship between the radio jet inclination angle  
\citep[taken from ][see Table \ref{tab:radio}]{Marin2016} and the average torus/accretion disk inclination angle. Our results indicate that while the jet falls inside the opening angle of the torus (for most of our sources), the orientation of these structures does not always align.}

\reff{At first glance, the lack of a statistically significant correlation between the parameters in Figure \ref{fig:rcd} seems to be inconsistent with the Unification model in which the flux of the core component decreases as the inclination angle increases. However, there are several factors contributing to these discrepancies. As noted in Section \ref{sec:3CRsed}, the shape of the IR SED (consequently the best-fit torus) is affected by the potential non-thermal component in some sources (e.g., 3C~43/212).
Additionally, we speculate that changes in accretion disk/torus orientation after jet launching may ruin any alignments between the two \maz{(although this may not lead to the systematic difference seen in Figure \ref{fig:rcd})}. 
Some studies find that a large misalignment between the jet and accretion disk ($>45^{\circ}$) may tear the disk apart into rings and lead to rapid accretion \citep[e.g.,][]{Nealon2015,Combes2021} and consequently a different SED. A more definitive study will be carried out when we include the radio galaxies, which will extend the range of orientation.}

We also examined the relation between $\textrm R_{CD}$ and all the other torus parameters (e.g., filling factor, the optical depth of the clouds or disk) and the accretion disk parameters and did not find any significant correlation. 
 Given the limited range of $\textrm R_{CD}$ of the quasars in our sample ($\sim 0-0.4$ without 3C 245), in order to see the variation of the radio core dominance and obscuration with the inclination angle, we need to include the edge-on sources. This will be addressed in a later paper, in which we will also compare our SED fits with the clear orientation-dependence of the X-ray obscuration in this sample \citep{Wilkes2013}.

\subsection{Comparisons with the Literature on the SED Fitting of 3CR Sample at \z}
\label{sec:3CR_lit}

The SED decomposition of 3CR radio sources at different redshifts and wavelengths has been the subject of several studies \citep[e.g.,][]{Drouart2014, Westhues2016,pece15,pece2016}, \reff{but these had mostly a limited wavelength range.} \pece\ investigated 1-1000\,\um SED of 3CR quasars and radio galaxies at \z (our parent sample) by fitting the AGN and host galaxy components simultaneously. To model the radiation from the torus, they use the  \cite{HK2010} clumpy torus model with a modified blackbody component to describe the emission from hot dust in the torus. To model the host galaxy component they use a modified blackbody at FIR wavelengths.

We compared the SEDs of all our sources with those from \pece\ to identify any discrepancies between the physical parameters obtained from the two fitting procedures. As noted in Section~\ref{sec:torus}, in the \sieb\ torus model, fluffy dust grains that are larger than the standard ISM are adopted which can survive closer to the BHs, 
resulting in stronger NIR radiation as well as more pronounced FIR/submm emission. \reff{The torus is the dominant source of radiation in our sources in the $\sim$ 2--30 \um regime, which roughly translates to 100--2000 K dust temperature, while the host galaxy peaks at $\sim$ 60--100 \um wavelengths (except for 3C 191/268.4) resulting in dust temperatures $\sim$ 25-50 K.  While the host galaxy dominates at FIR wavelengths, the composition of the dust grains and the clumpy structure in our torus model results in more contribution from the torus at FIR wavelengths than \pece. This is particularly noticeable in 3C\,190
resulting in a significantly lower estimated SFR (from 470 $\rm M_{\odot}/yr$ in \citeauthor{pece15} to 257 $\rm M_{\odot}/yr$ from ARXSED. However, this is not the case for 3C\,191, and 3C\,268.4 in which the torus is dominant at FIR wavelengths. 
}

Also, \pece\ considered the host galaxy emission only at FIR wavelengths while our host galaxy model covers UV to radio wavelengths (although it is not the dominant source of radiation in most of these bands). Another difference is the way upper limits are treated in the two models. \citeauthor{pece15} treated the upper limits (especially at 70, 160, and 250\,$\mu$m) as detections during the fit and report the SFR estimated from these bands as upper limits, while we do treat them as upper limits. Finally, unlike ARXSED, \pece\ did not consider non-thermal radiation from the radio structures in the submm/FIR bands. Overall, these differences result in lower SFR estimates from our fitting procedure. 
\reff{We will discuss the host galaxy properties (e.g., SFR, stellar mass) and  discrepancies with \pece\ fully in a future paper}. 

\cite{pece_sed2} improved upon the approach of \pece\ by adopting the \sieb\ torus model and the P{\'E}GASE SED model \citep{Fioc2019} for the host galaxies, while still limited to 1--1000\um. However, their improved SED model, was only run for 12 NLRG of the \pece\ sample sources \reff{and therefore are directly not comparable to our sources.}

\subsection{Non-thermal Radiation at Observed mm to FIR Wavelengths} \label{sec:non_th}
\begin{figure}[t] 
    \includegraphics[width=0.4\textwidth,angle =90]{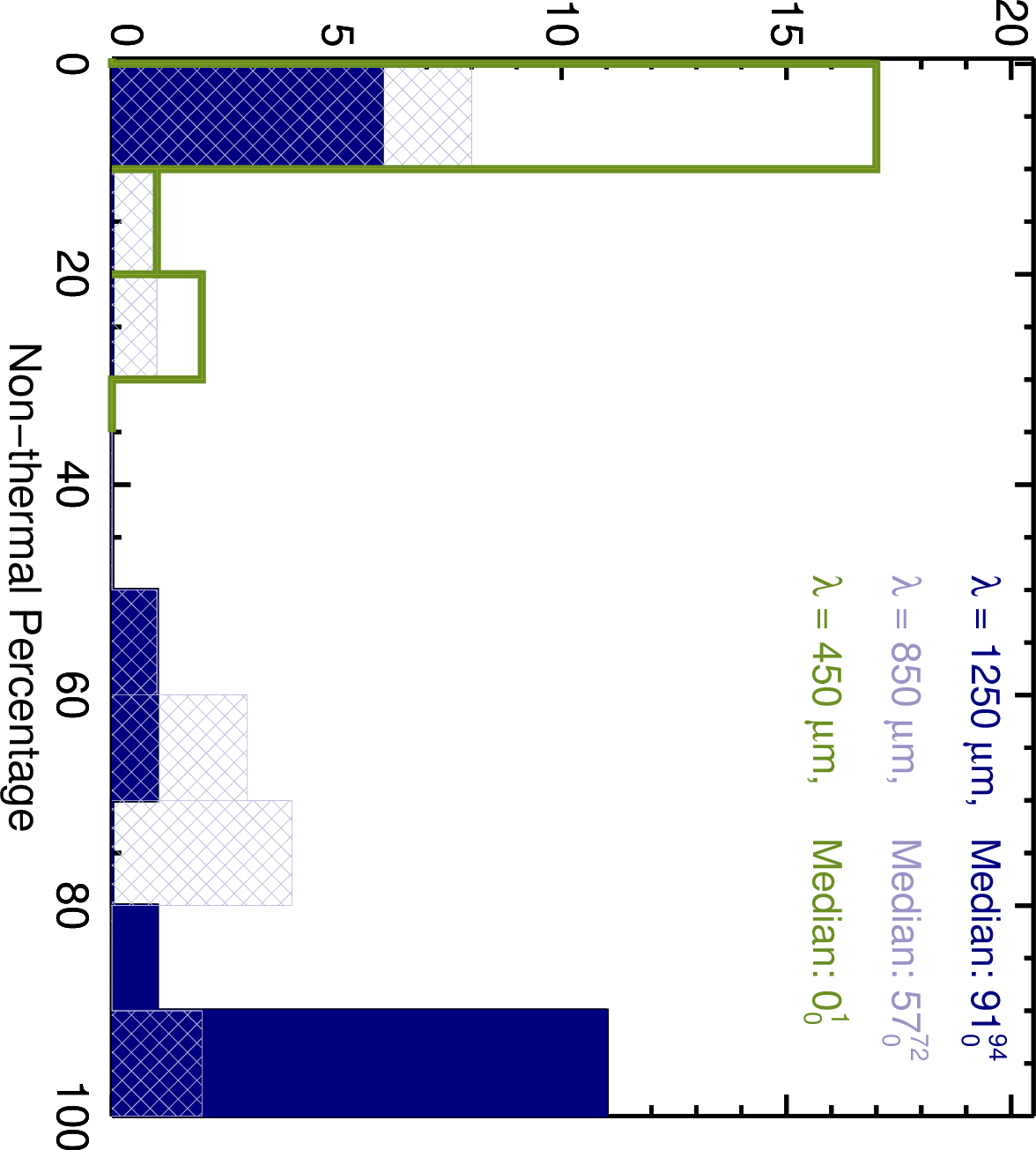}
             \caption{\reff{The percentages of non-thermal emission originating from the radio structures at different observed wavelengths. At 1250 \um, the radio component is dominant in the majority of our sources, while at FIR, a combination of the host galaxy and torus dominates. At 850 \um, the thermal emission from dust and non-thermal emission from the radio structures have comparable contributions.}}
\label{fig:nth}
\end{figure}

\reff{Figure \ref{fig:nth} illustrates the fraction of non-thermal radiation in our sample calculated at mm to FIR observed wavelengths. The best-fit radio model, along with the host galaxy and the torus, is the main component contributing to these wavelengths. Figure \ref{fig:nth} shows at 1.25 mm (observed-frame), the total emission is dominated by the radio structure, while at FIR, the torus and the host galaxies are dominant. Our model predicts six sources (out of 20) have minimal synchrotron contribution at 1.25 mm. These are the quasars with a radio emission cutoff that occurs at wavelengths longer than 1.25 mm. Figure \ref{fig:nth} also shows that the emission from dust (host galaxy and torus) dominates the synchrotron emission at 450\um (observed-frame). However, in three of our sources (3C~190/191/268.4), there remains $\sim$20\% contribution from the radio structures at 450 \um.}

\reff{At 850 \um (observed-frame), the thermal emission from the dust and non-thermal emissions from the radio structures are comparable. This implies the submm-based SFR in radio-loud quasars should be calculated after removing potential contamination from radio structures. While in a few of our sources (i.e., 3C 212/245), the 850 \um photometry is dominated by the core contribution, in others the source of the emission,  whether from extended structures/lobes or the interaction of the radio structures with the star-forming gas, is not clear. Further observations with a mm-wave interferometer are required to understand the nature of submm radiation in these sources.}

In addition, there are two sources in \reff{Figure \ref{fig:nth} (3C~14/186) with comparable thermal 
and non-thermal emission 
at 1.25 mm. While 3C~14 has the most extended jet in our sample, 3C~186 is a CSS. However, the shape of the radio spectrum of 3C~186 may suggest that it is confined by a dense environment rather than being young \cite[e.g.,][]{Fanti1995,Hilbert2016}.}

\reff{\cite{Haas2006} observed seven quasars from our sample (along with 4 other sources) and found evidence for the synchrotron nature of the observed 1.25 mm and 850 \um radiation. Consistent with \cite{Haas2006}, our findings do not confirm earlier results 
\citep[e.g.,][]{Willott2002} that found the 850 \um emission of quasars at $z\sim1.5$ is dominated by the thermal emission from dust.}

\begin{figure*}[th!]
        \includegraphics[width=0.27\textwidth,angle =90]{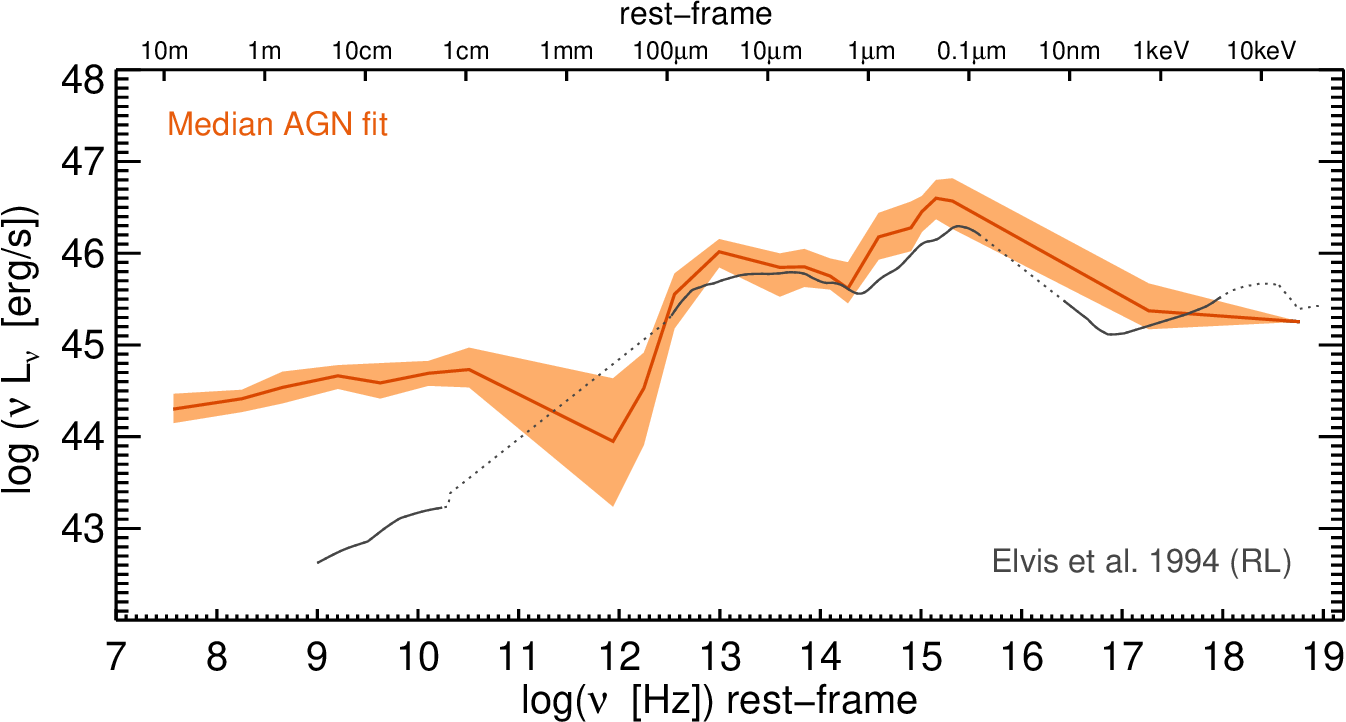} 
    \includegraphics[width=0.27\textwidth,angle =90]{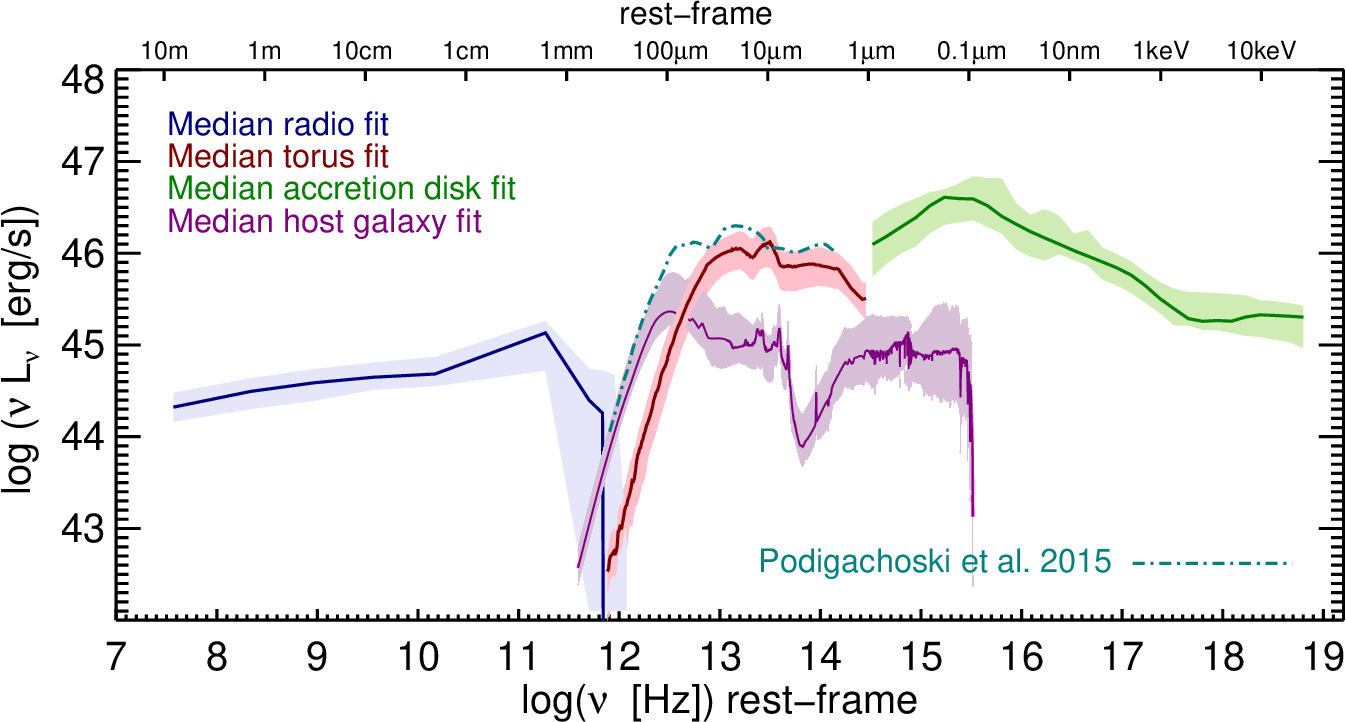}
             \caption{\textit{Left:} The median SED of the AGN fits in our sample. The gray curve shows the median SED of the radio-loud quasars in \cite{Elvis1994} normalized at 1.5\,$\mu$m. The dotted gray curve indicates regions with few or no data available in the \citeauthor{Elvis1994} sample.
             \textit{Right:} The median SED of each of the components used in this study. The dotted-dashed curve is the median SED of \pece\ which includes the AGN and host galaxy components combined.
            The shaded region around each component indicates the 25th-75th percentile ranges. The small number of photometric measurements between the radio and FIR bands results in a larger scatter around the median in these bands.}    
\label{fig:medsed}
\end{figure*}

\subsection{The Median SED of 3CRR Quasars}
\label{sec:medsed}

 Figure \ref{fig:medsed} presents the median AGN SED  (left) and the medians for the individual components (right) in our study. The shaded region around each component shows 25th-75th percentile ranges. Because the number of measurements between the radio and FIR is small, the scatter around the median is significantly larger in these bands.
 The gray curve (in the left panel) indicates the median SED of the radio-loud quasars in \cite{Elvis1994} normalized at 1.5\,\um and the dotted parts indicate regions with few or no data available in \cite{Elvis1994} sample. The dotted-dashed curve (in the right panel) is the median SED of \pece. The median SED of \cite{Elvis1994} was obtained from 18 X-ray-bright radio-loud quasars at $z<0.8$ after subtracting the host galaxy contribution. The median SED of \pece\ was obtained from a sample of 25 3CR radio-loud quasars at \z (our parent sample) using the SED components described in Section \ref{sec:3CR_lit}. 

The median SED of the \cite{Elvis1994} radio-loud quasars covers a similarly broad frequency range and is \reff{qualitatively} in agreement with our median SED. However, there are differences between the two curves at some specific bands. \citeauthor{Elvis1994}'s sample has a lower redshift, given the nature of their selection (X-ray bright), but they can still be as luminous as our quasars at X-ray wavelengths. \maz{The slope of the X-ray power-law in  our sample is softer than 
\citeauthor{Elvis1994}, which could be either due to the way the two samples are selected (radio-selected versus X-ray selected) and/or the presence of a stronger radio-linked X-ray component in our sample \citep[see also][]{shang2011}. Indeed, \citeauthor{Elvis1994}'s sources, despite being radio-loud, are much less luminous than our sources at radio bands.}
\citeauthor{Elvis1994}'s IR observations were limited to {\sl IRAS} wavelengths, while ours extend to longer
wavelengths and 
\reff{our sources, amongst the most luminous radio sources known, are $\sim1.7$ times more luminous at 
1--100 \um .}

\reff{Another noticeable difference is the much steeper FIR/submm slope 
(in $ \rm log(\nu) -log (\nu L_{\nu})$ space). Indeed the \citeauthor{Elvis1994} sample lacks the data required to constrain the SED at these wavelengths \citep[also see][]{shang2011}.}
Overall, considering the differences at radio and visible-UV wavelengths, the \citeauthor{Elvis1994} median SED does not represent the average behavior of luminous radio-selected AGN at $z>1$. 

Figure \ref{fig:medsed} (right) shows the median SED of the quasars (and their host galaxies) from \pece. Given the scatter in the \citeauthor{pece15} SEDs (0.5-1 dex, not shown for clarity), their median SED is generally consistent with ours.
One notable difference is the redder and brighter FIR peak relative to our sample. This may be due to the difference in the torus models, i.e., the larger grains adopted in ARXSED, which results in stronger FIR radiation from the torus and consequently less pronounced emission from the host galaxy. We also note that ARXSED subtracts the non-thermal radiation from radio structures in the submm/FIR, which can contribute to a fainter FIR peak. 

There have been many attempts to derive the average SED of quasars \citep[e.g.][among others]{Richards2006, shang2011}. 
\reff{\cite{Richards2006} studied a sample of SDSS quasars, mostly from FIR-to-X-ray (with an additional eight radio-loud quasars) data, and presented their median SEDs. At X-ray bands, our sources are brighter than \cite{Richards2006}, which could be either due to the difference between the SDSS and 3CRR sample or could be attributed to a stronger radio-linked X-ray emission in our sample. At submm/FIR wavelengths, \citeauthor{Richards2006} SEDs have a flatter slope ($ \rm log(\nu) -log (\nu L_{\nu})$ space) than ours (and \citealt{Elvis1994}) possibly due to the lack of data at these wavelengths, or the smaller number of sources.} 

\reff {\cite{shang2011} presented the median radio to X-rays SED of a sample of 85 (27 radio-quiet and 58 radio-loud) optically bright, non-blazar quasars. Rather than using detailed SED models, \cite{shang2011} used H-band photometry as a proxy of the host galaxy contribution and corrected the photometry at other wavelengths accordingly. At X-rays, our findings are overall consistent (given that their sample includes luminous radio sources at comparable redshifts). \citeauthor{shang2011} found more variation in the SED than \citeauthor{Elvis1994} and \citeauthor{Richards2006} by including FIR observations.} 

\reff{In summary, our findings are qualitatively consistent with the literature. However, we provide more detailed features in the shape of the median SED by including Herschel and mm/submm photometry. Additionally, we present the median SED using the best-fit model for each component derived using 
state-of-the-art models in the literature (i.e., torus, accretion disk). To obtain the various components we tie them to each other in a physically meaningful way and compare our fits with the observations at different wavelengths for individual sources as a sanity check. Another advantage of our median SED is detailed modeling of the host galaxy emission which all the studies above lack.}

%


\section{Summary} 
\label{sec:summary}

In this study, we present a state-of-the-art AGN radio-to-X-ray SED model (ARXSED) that simultaneously fits AGN and the host galaxy components. Fitting at radio wavelengths requires four models to account for radiation from the lobes as well as compact radio structures such as radio cores and hot spots (where a superposition of multiple self-absorbed components makes the shape of the spectrum more complex). Additionally, the model accounts for a steepening or cut-off due to the aging of the electron populations. Emission from the torus is fitted by the two-phase torus model of  \cite{Siebenmorgen2015} in which the dust can be distributed in a homogeneous disk,
a clumpy medium, or a combination of both. The visible-UV-X-ray radiation from the accretion disk is fitted with the QSOSED model developed by  \cite{Kubota2018}. The emission from the host galaxy is fitted using an underlying component from UV to radio wavelengths \citep[MAGPHYS][]{dc2008,dcr2015}.

Using ARXSED, we fit the radio-to-X-ray SED of a sample of 20 radio-loud quasars from the 3CRR sample at \z. The 3CRR sample includes low-frequency radio-selected AGN and so is unbiased in terms of orientation and dust obscuration. \maz{Hence the distribution of their orientation angles can be predicted.}
We have compiled the SED by combining archival multi-frequency radio observations, recent SMA/ALMA observations, \textit{Herschel}, \textit{WISE}, \textit{Spitzer}, {2MASS}, {UKIRT}, {SDSS}, \textit{XMM-Newton} and \textit{Chandra} for our analysis (Section \ref{sec:data}). In order to obtain the intrinsic SED of the AGN in our sample, we correct the photometry for 
the reddening and absorption in the host galaxy, the Milky Way, as well as the dusty torus (Section \ref{sec:dered}). Our SED models successfully reproduce the observed photometry and constrain the parameters describing the structures surrounding SMBHs at $z>1$.  Given the large number of parameters, diverse data quality, and possible variability, which may bias our SED fits, we apply priors to the fits based on independent measurements of parameters such as $M_{BH}$ from the literature. This helps to ensure consistency of the SED fitting results, which otherwise may not be unique.

 In this paper, we present the fitting results for individual sources (Section \ref{sec:quasar_sed}) and the physical properties of the AGN components derived from our modeling (Section \ref{sec:3CRsed}). 
Our main findings are as follows:

\begin{itemize}
   
\item A simple power-law ($L_{\nu}\propto \nu^{\alpha}$), is unable to replicate the radio emission from our sources when complex radio structures (i.e., lobes, jets, cores, hot spots) are present (Section \ref{sec:quasar_sed}).

\item 
\reff{ARXSED predicts that non-thermal emission from the radio structures contributes  91-57\% 
to the 3CRR quasars SED between 1.25 mm and 850\,$\mu$m.} It is important to subtract this source of contamination when broadband photometry is used to estimate host galaxy properties such as dust mass and star formation rate (Section \ref{sec:non_th}).

\item The median properties of the best-fit torus parameters and their associated 25th-75th percentile ranges are: 
the inner radius of the torus, $\rm R_{in}= 2.8_{1.8}^{4.5}$ (pc), 
the filling factor of the clumps, $\rm V_{C}= 1.5_{1.5}^{7.7}$ (\%),
the optical depth of the clumps, $\rm \tau_{C}=4.5_{0.0}^{45.0}$,
the optical depth of the homogeneous disk,
$\rm \tau_{D}=300_{300}^{1000}$ (Section \ref{sec:3CR_lit}).

\item The median properties of the best-fit accretion disk parameters and their associated 25th-75th percentile ranges are as follows:  mass of the BH, $\rm M_{BH}/M_{\sun}=  2.9_{1.7}^{6.0} (\times 10^9)$, the logarithm of Eddington ratio $\log (\lambda_{Edd})=  -1.0_{-1.2}^{-0.6} $,
the dimensionless spin parameter $\rm a(\equiv Jc/GM_{BH}^{2}$) $=  0.98_{0.94}^{0.99} $. The SMBH properties estimated by ARXSED agree with those \reff{extracted using  emission line techniques} in the literature for similarly defined samples (Section \ref{sec:3CR_lit}).

\item 
\reff{We find the median (and associated 25th-75th percentile range) inclination angle of 
$52_{47}^{57} {}^\circ$
from the best-fit torus/accretion disk, while the average inclination angle of the radio jets reported in the literature for our sample is $33^{\circ}\pm14^{\circ}$. We find that the inclination angle from the torus/accretion disk is not the same as the radio jet. This could be due to an underlying non-thermal contribution from radio structures that changes the shape of the SEDs at shorter wavelengths (affecting the best-fit torus and accretion disk models) and/or the presence of red quasars in our sample. It is also plausible that this misalignment occurs after the jets are launched (Section \ref{sec:orientation}). Additionally, we note that the quasars in our sample have a limited range of radio core dominance, $\rm R_{CD}$ (and jet inclination). To investigate the relation of the jet inclination with the torus/accretion disk parameters, edge-on sources with a wider range of $\rm R_{CD}$ should be considered as well (Section \ref{sec:orientation}).}

\item We present the median intrinsic SED of the radio-loud quasars at \z. \reff{Our median SED self-consistently covers the gap in observations  and provides more detailed features in the shape of the SED compared to the literature.} We find that the median SED of \cite{Elvis1994}, obtained based on a sample of radio-loud quasars at $z<1$, does not describe the SEDs of luminous, radio-selected AGN at $z>1$. The difference between the two median SEDs could be due to sample selection, redshift/luminosity, and/or observation limitations (Section \ref{sec:medsed}).

\end{itemize}

\section*{Acknowledgements} 

We thank the referee for their positive comments and constructive advice, which helped improving the paper. Support for this work was provided by the National Aeronautics and Space Administration and the \textit{Chandra} X-ray Center (CXC), which is operated by the Smithsonian Astrophysical Observatory and on behalf of the
National Aeronautics Space Administration under contract NAS8-03060
(BJW, MAz,JK). Additional support was provided by NASA NuSTAR grant \#80NSSC21K0058 (JK).%

The scientific results in this article are based significantly on observations made by the \textit{Chandra}~X-ray Observatory (CXO).
This research has made use of data obtained from the \textit{Chandra} Data
Archive.

This research has made use of data provided by the National Radio
Astronomy Observatory which is a facility of the National Science
Foundation operated under a cooperative agreement with Associated
Universities, Inc., and data from the Sloan Digital Sky Survey (SDSS). 
Funding for the SDSS and SDSS-II has been provided by the Alfred
P. Sloan Foundation, the Participating Institutions, the National
Science Foundation, the U.S. Department of Energy, the National
Aeronautics and Space Administration, the Japanese Monbukagakusho, the
Max Planck Society and the Higher Education Funding Council for
England. The SDSS Web Site is http://www.sdss.org/. 
The SDSS is managed by the Astrophysical Research Consortium for the
Participating Institutions. The Participating Institutions are the
American Museum of Natural History, Astrophysical Institute Potsdam,
University of Basel, University of Cambridge, Case Western Reserve
University, University of Chicago, Drexel University, Fermilab, the
Institute for Advanced Study, the Japan Participation Group, Johns
Hopkins University, the Joint Institute for Nuclear Astrophysics, the
Kavli Institute for Particle Astrophysics and Cosmology, the Korean
Scientist Group, the Chinese Academy of Sciences (LAMOST), Los Alamos
National Laboratory, the Max-Planck-Institute for Astronomy (MPIA),
the Max-Planck-Institute for Astrophysics (MPA), New Mexico State
University, Ohio State University, University of Pittsburgh,
University of Portsmouth, Princeton University, the United States
Naval Observatory, and the University of Washington.

This research is based on observations made by {\it Herschel}, which
is an ESA space observatory with science instruments provided by
European-led Principal Investigator consortia and with important
participation from NASA.
This work is based in part on observations made with the Spitzer Space
Telescope, which was operated by the Jet Propulsion Laboratory,
California Institute of Technology under a contract with NASA.

We acknowledge the use of Ned Wright's calculator
\citep{2006PASP..118.1711W} and NASA/IPAC Extragalactic Database
(NED), operated by the Jet Propulsion Laboratory, California Institute
of Technology, under contract with the National Aeronautics and Space
Administration.

We acknowledge the use of IRAF, which is distributed by the National Optical Astronomy Observatories, and operated by the Association of Universities for Research in Astronomy, Inc., under contract to the National Science Foundation.

\bibliographystyle{apjurl}
\bibliography{references.bib}

\appendix 
Tables A1, A2 and A3 below present high, medium, and low-frequency radio data used in the SED analysis in this study with their references. The flux densities are in Jansky. 

\begin{sidewaystable}[t]
\vspace*{-5cm}
\tablenum{1}
\caption{High frequency radio data}
\begin{minipage}{\textwidth}

\tiny{
   \begin{tabular}{lccccccccccccccccccccccccccccccccc}  
\hline
Name&230 GHz&90 GHz&31.4 GHz&14900 MHz&10700 MHz&8870 MHz&8400 MHz&8085 MHz&5000 MHz&4.85 GHz&2695 MHz
\\
\hline \hline
\\
3C 9&&&&0.13  $\pm$ 0.01&0.22  $\pm$ 0.02&&&&0.55  $\pm$ 0.05&0.48  $\pm$ 0.07&0.98  $\pm$ 0.04\\
&&&&L80&L80&&&&L80&G91&L80\\
3C 14&&&&0.15  $\pm$ 0.01&0.22  $\pm$ 0.02&&&&0.61  $\pm$ 0.03&0.55  $\pm$ 0.07&1.02  $\pm$ 0.04\\
&&&&L80&L80&&&&L80&G91&L80\\
3C 43&0.06  $\pm$ 0.01&0.19  $\pm$ 0.04&&0.45  $\pm$ 0.01&0.61  $\pm$ 0.04&&&&1.08  $\pm$ 0.04&1.17  $\pm$ 0.16&1.70  $\pm$ 0.04\\
&SJ95&SJ95&&L80&L80&&&&L80&G91&L80\\
3C 181&&&&0.20  $\pm$ 0.01&0.29  $\pm$ 0.02&&&&0.66  $\pm$ 0.05&0.68  $\pm$ 0.09&1.25  $\pm$ 0.04\\
&&&&L80&L80&&&&L80&G91&L80\\
3C 186&&&&&0.08  $\pm$ 0.01&&&&0.38  $\pm$ 0.05&0.31  $\pm$ 0.04&0.59  $\pm$ 0.04\\
&&&&&L80&&&&L80&G91&L80\\
3C 190&&&&0.30  $\pm$ 0.01&0.36  $\pm$ 0.02&&&&0.81  $\pm$ 0.06&0.7  $\pm$ 0.1&1.40  $\pm$ 0.05\\
&&&&L80&L80&&&&L80&G91&L80\\
3C 191&&&&0.18  $\pm$ 0.01&0.23  $\pm$ 0.02&&&&0.46  $\pm$ 0.06&0.56  $\pm$ 0.08&0.96  $\pm$ 0.04\\
&&&&L80&L80&&&&L80&G91&L80\\
3C 204&&&&0.11  $\pm$ 0.01&0.16  $\pm$ 0.01&&&&0.34  $\pm$ 0.03&0.37  $\pm$ 0.03&0.51  $\pm$ 0.04\\
&&&&L80&L80&&&&L80&G91&L80\\
3C 205&&&&0.16  $\pm$ 0.01&0.23  $\pm$ 0.02&&&&0.67  $\pm$ 0.04&0.69  $\pm$ 0.07&1.12  $\pm$ 0.04\\
&&&&L80&L80&&&&L80&G91&L80\\
3C 208&&&&0.14  $\pm$ 0.01&0.23  $\pm$ 0.02&&&&0.54  $\pm$ 0.05&0.51  $\pm$ 0.07&0.96  $\pm$ 0.04\\
&&&&L80&L80&&&&L80&G91&L80\\
3C 212&&&&0.48  $\pm$ 0.01&0.50  $\pm$ 0.02&&&&0.88  $\pm$ 0.04&0.7  $\pm$ 0.1&1.43  $\pm$ 0.04\\
&&&&L80&L80&&&&L80&G91&L80\\
3C 245&&&0.94  $\pm$ 0.15&0.85  $\pm$ 0.01&0.97  $\pm$ 0.04&1.0  $\pm$ 0.0&&&1.38  $\pm$ 0.04&1.74  $\pm$ 0.24&2.0  $\pm$ 0.0\\
&&&G81&L80&L80&S73&&&L80&G91&W75\\
3C 268.4&&&&0.18  $\pm$ 0.01&0.24  $\pm$ 0.02&&&&0.60  $\pm$ 0.03&0.61  $\pm$ 0.07&1.07  $\pm$ 0.04\\
&&&&L80&L80&&&&L80&G91&L80\\
3C 270.1&&&&0.44  $\pm$ 0.01&0.45  $\pm$ 0.02&&&&0.86  $\pm$ 0.04&0.88  $\pm$ 0.11&1.51  $\pm$ 0.04\\
&&&&L80&L80&&&&L80&G91&L80\\
3C 287&0.12  $\pm$ 0.02&0.51  $\pm$ 0.09&&1.42  $\pm$ 0.03&1.73  $\pm$ 0.05&&&2.20  $\pm$ 0.18&3.24  $\pm$ 0.06&3.11  $\pm$ 0.41&4.65  $\pm$ 0.06\\
&SJ95&SJ95&&L80&L80&&&C83&L80&G91&L80\\
3C 318&&&&0.23  $\pm$ 0.01&0.32  $\pm$ 0.02&&&&0.74  $\pm$ 0.03&0.85  $\pm$ 0.12&1.34  $\pm$ 0.04\\
&&&&L80&L80&&&&L80&G91&L80\\
3C 325&&&&0.25  $\pm$ 0.02&0.39  $\pm$ 0.02&&&&0.82  $\pm$ 0.04&1.1  $\pm$ 0.1&1.86  $\pm$ 0.04\\
&&&&L80&L80&&&&L80&G91&L80\\
4C 16.49&&&&&&&&&0.32  $\pm$ 0.0&0.34  $\pm$ 0.05&0.68  $\pm$ 0.0\\
&&&&&&&&&W90&G91&W90\\
3C 432&&&&0.08  $\pm$ 0.01&0.13  $\pm$ 0.01&&&&0.31  $\pm$ 0.06&0.41  $\pm$ 0.06&0.77  $\pm$ 0.04\\
&&&&L80&L80&&&&L80&G91&L80\\
3C 454.0&&&&0.30  $\pm$ 0.01&0.39  $\pm$ 0.03&&&&0.78  $\pm$ 0.03&0.79  $\pm$ 0.11&1.22  $\pm$ 0.04\\
&&&&L80&L80&&&&L80&G91&L80&\\
\hline
\end{tabular}

}
\end{minipage}
\end{sidewaystable}

\begin{sidewaystable}[t]
\vspace*{-5cm}
\tablenum{2}
\caption{Medium frequency radio data}
\begin{minipage}{\textwidth}
\tiny{
   \begin{tabular}{lccccccccccccccccccccccccccccccccc}
\hline
Name&1400 MHz&750 MHz&635 MHz&408 MHz&365 MHz&326MHz&178 MHz&160 MHz&151 MHz&86 MHz\\
\hline\hline
\\
3C 9&1.96  $\pm$ 0.12&3.89  $\pm$ 0.19&&7.79  $\pm$ 0.32&&&&&&35.7 $\pm$ 1.6\\
&L80&L80&&L81&&&&&&L80\\
3C 14&1.9  $\pm$ 0.1&3.58  $\pm$ 0.21&&5.85  $\pm$ 0.18&6.83  $\pm$ 0.09&&11.3 $\pm$ 1.1&&&24.4 $\pm$ 2.7\\
&L80&L80&&L81&D96&&L80&&&L80\\
3C 43&2.82  $\pm$ 0.08&4.28  $\pm$ 0.18&&6.44  $\pm$ 0.52&8.56  $\pm$ 0.08&&12.6 $\pm$ 1.3&16.3  $\pm$ 2.4&&20.5 $\pm$ 4.4\\
&L80&L80&&F74&D96&&L80&K81&&L80\\
3C 181&2.30 $\pm$ 0.69&3.76  $\pm$ 0.17&&6.73  $\pm$ 0.21&7.62  $\pm$ 0.16&&15.80 $\pm$ 0.79&&&24.2 $\pm$ 2.2\\
&C98&L80&&L81&D96&&L80&&&L80\\
3C 186&1.24 $\pm$ 0.04&2.74  $\pm$ 0.17&&5.55  $\pm$ 0.11&6.58  $\pm$ 0.11&&15.37 $\pm$ 0.77&&15.59  $\pm$ 0.07&33.2 $\pm$ 2.7\\
&C98&L80&&F85&D96&&L80&&H93&L80\\
3C 190&2.55  $\pm$ 0.07&4.30  $\pm$ 0.17&&7.63  $\pm$ 0.24&9.09  $\pm$ 0.09&&16.35 $\pm$ 0.82&&&26.8 $\pm$ 2.7\\
&L80&L80&&L81&D96&&L80&&&L80\\
3C 191&1.85 $\pm$ 0.06&3.44  $\pm$ 0.17&&7.32  $\pm$ 0.32&7.49  $\pm$ 0.08&&14.17 $\pm$ 0.71&&&34.3 $\pm$ 1.6\\
&C98&L80&&L81&D96&&L80&&&L80\\
3C 204 &1.30  $\pm$ 0.08 &2.37  $\pm$ 0.06 &&7.32  $\pm$ 0.32&5.49 $\pm$ 0.17&&11.45 $\pm$ 0.57&& 
&29.6 $\pm$ 1.1\\ 
&L80&P66&&L81&D96&7.9&L80&&&L80\\
3C 205&2.39  $\pm$ 0.07&3.95  $\pm$ 0.18&&&9.3  $\pm$ 0.2&&13.7 $\pm$ 1.4&&&39.2 $\pm$ 4.4\\
&L80&L80&&&D96&&L80&&&L80\\
3C 208&2.36 $\pm$ 0.07&4.56  $\pm$ 0.17&&7.75  $\pm$ 0.25&10.33 $\pm$ 0.18&&20.0 $\pm$ 1.0&&&41.5 $\pm$ 2.4\\
&C98&L80&&L81&D96&&L80&&&L80\\
3C 212&2.37  $\pm$ 0.07&4.44  $\pm$ 0.18&&7.01  $\pm$ 0.22&8.34  $\pm$ 0.15&&16.46  $\pm$ 0.82&&&29.8 $\pm$ 3.1\\
&C98&L80&&L81&D96&&L80&&&L80\\
3C 245&3.3 $\pm$ 0.1&5.06  $\pm$ 0.18&&8.90  $\pm$ 0.37&9.45  $\pm$ 0.21&&15.70  $\pm$ 0.79&&&27.4  $\pm$ 1.1\\
&C98&L80&&L81&D96&&L80&&&L80\\
3C 268.4&1.98 $\pm$ 0.06&3.62  $\pm$ 0.17&&5.69  $\pm$ 0.12&6.58  $\pm$ 0.12&&11.23  $\pm$ 0.56&&&22.4  $\pm$ 2.7\\
&C98&L80&&F85&D96&&L80&&&L80\\
3C 270.1&2.73  $\pm$ 0.07&5.0  $\pm$ 0.1&&8.20  $\pm$ 0.41&9.74  $\pm$ 0.13&&14.82  $\pm$ 0.74&&&30.8  $\pm$ 2.7\\
&L80&P66&&C70&D96&&L80&&&L80\\
3C 287&7.05 $\pm$ 0.21&9.67  $\pm$ 0.25&&11.94 $\pm$ 0.95&14.96 $\pm$ 0.28&&17.44 $\pm$ 0.87&20  $\pm$ 3&15.5  $\pm$ 0.7&19.4 $\pm$ 2.2\\
&C90&L80&&C72&D96&&L80&K81&W96&L80\\
3C 318&2.56  $\pm$ 0.07&2.69 $\pm$ 0.08&4.44  $\pm$ 0.18&&&9.21  $\pm$ 0.09&&13.41  $\pm$ 0.67&&13.98  $\pm$ 0.64\\
&L80&C98&L80&&&D96&&L80&&W96\\
3C 325&3.56  $\pm$ 0.13&6.19  $\pm$ 0.19&&&12.12 $\pm$ 0.29&&17.0 $\pm$ 0.9&&&33.1 $\pm$ 2.7\\
&C98&L80&&&D96&&L80&&&L80\\
4C 16.49&1.46  $\pm$ 0.05&&&5.36  $\pm$ 0.17&6.3  $\pm$ 0.1&&10.50  $\pm$ 0.84&&&\\
&C98&&&L81&D96&&G67&&&\\
3C 432&1.58 $\pm$ 0.06&2.90  $\pm$ 0.17&&6.05  $\pm$ 0.15&6.48  $\pm$ 0.12&&12.0  $\pm$ 1.2&&&21.1  $\pm$ 1.9\\
&C98&L80&&L81&D96&&L80&&&L80\\
3C 454.0&2.14  $\pm$ 0.07&3.46  $\pm$ 0.17&&5.65  $\pm$ 0.18&6.49  $\pm$ 0.05&&12.64  $\pm$ 0.63&&&\\
&L80 &L80 &&L81 &D96 &&L80 &&&&\\           
\hline
\end{tabular}

}
\end{minipage}
\end{sidewaystable}

\begin{sidewaystable}[t]
\vspace*{-4cm}
\tablenum{3}

\caption{Low frequency radio data}
\begin{minipage}{\textwidth}
\tiny{
   \begin{tabular}{lccccccccccccccccccccccccccccccccc}
\hline
Name&80MHz&74 MHz&60 MHz&38 MHz&26.3 MHz&25.0MHz&22.25 MHz&20.0 MHz&16.7 MHz&14.7 MHz&12.6 MHz&10 MHz\\
\hline\hline
\\
3C 9&&&53.0  $\pm$ 7.1&75.5 $\pm$ 7.6&99.0 $\pm$ 18.0&&97.8 $\pm$ 16.7&&&&&204.0$\pm$ 42.0\\
&&&A68&L80&L80&&L80&&&&&L80\\
3C 14&&&&42.5 $\pm$ 8.5&69.0 $\pm$ 7.0&&&&&&&\\
&&&&L80&L80&&&&&&&\\
3C 43&23.0  $\pm$ 3.0&&&46.0 $\pm$ 9.2&46.0 $\pm$ 6.0&&69.0 $\pm$ 8.3&&&&&\\
&K81&&&L80&L80&&L80&&&&&\\
3C 181&&&54.0   $\pm$4.0&37.8 $\pm$ 7.6&57.0 $\pm$ 8.0&&&&&&&\\
&&&A68&L80&L80&&&&&&&\\
3C 186&&&42.02 $\pm$ 3.01&57.8 $\pm$ 5.8&54.0 $\pm$ 7.0&77.0  $\pm$ 24.0&54.1  $\pm$ 5.0&61.0  $\pm$ 17.0&103.0 $\pm$ 35.0&126.0  $\pm$ 45.0&120.0  $\pm$ 47.0&\\
&&&A68&L80&L80&B70&L80&B70&B70&B70&B70&\\
3C 190&&&40.1 $\pm$ 8.0&64.0  $\pm$ 8.0&&71.3 $\pm$ 11.7&&&&&&192.0 $\pm$ 42.0\\
&&&&L80&V75&&L80&&&&&L80\\
3C 191&&&&56.6 $\pm$ 5.7&46.0  $\pm$ 8.0&&96.6 $\pm$ 11.7&120.0  $\pm$ 18.0&160.0$\pm$ 32.0&300.0$\pm$ 66.0&380.0$\pm$ 84.0&276.0$\pm$ 66.0\\
&&&&L80&L80&&L80&B69&B69&B69&B69&L80\\
3C 204& &26.6 $\pm$ 2.7&34.0 $\pm$3.0 &58.9 $\pm$ 10.0 
&&101.0 $\pm$ 19.2 && 95.0  $\pm$ 19.0& 131.0$\pm$ 26.2
& &380.0  $\pm$ 84.0&\\
& &C07&A68&H95& &B70
&&B70&B70&&B70&\\
3C 205&&&&54.3 $\pm$ 5.4&60.0 $\pm$ 10.0&&&88.0$\pm$ 16.7&85.0 $\pm$ 16.2&130.0  $\pm$ 28.6&147.0 $\pm$ 32.3&162.0  $\pm$ 36.0\\
&&&&L80&L80&&&B70&B70&B70&B70&L80\\
3C 208&&&42.02$\pm$ 3.01&68.44 $\pm$ 6.84&83.0 $\pm$ 10.0&&67.9 $\pm$ 21.7&&&&&\\
&&&A68&L80&L80&&L80&&&&&\\
3C 212&&&46.02$\pm$ 4.0 &&&&29.9  $\pm$ 11.7&&&&&\\
&&&A68&&&&L80&&&&&\\
3C 245&&&34.0$\pm$    5.0&47.2  $\pm$ 4.7&73.0  $\pm$ 8.0&&64.4 $\pm$ 11.7&137.0  $\pm$ 57.5&125.0  $\pm$ 35.0&132.0 $\pm$ 27.7&161.0  $\pm$ 45.1&170.4  $\pm$ 39.6\\
&&&A68&L80&L80&&L80&B70&B70&B70&B70&L80\\
3C 268.4&&20.7  $\pm$ 3.8&&36.6 $\pm$ 7.3&34.0 $\pm$ 8.0&&&&&&&\\
&&M05&&L80&L80&&&&&&&\\
3C 270.1&&30.8  $\pm$ 0.0&&33.0  $\pm$ 6.6&60.0  $\pm$ 8.0&&60.9  $\pm$ 10.0&85.0  $\pm$ 30.0&141.0 $\pm$ 65.0&210.0$\pm$ 61.0&&\\
&&C04&&L80&L80&&L80&B70&B70&B70&&\\
3C 287&&&&30.7 $\pm$ 6.1&&65.0  $\pm$ 21.0&&120.0 $\pm$ 50.4&140.0$\pm$ 61.6&240.0$\pm$ 120.0&320.0$\pm$ 160.0&\\
&&&&L80&&B69&&B69&B69&B69&B69&\\
3C 318&&&&&25.9$\pm$ 7.8&31.0 $\pm$ 7.0&&37.9  $\pm$ 3.3&&&&\\
&&&&&L80&L80&&L80&&&&\\
3C 325&&&&42.5 $\pm$ 4.3&&&&&&&&\\
&&&&L80&&&&&&&&\\
4C 16.49&&&&&42.0  $\pm$ 9.0&&&&&&&\\
&&&&&V75&&&&&&&\\
3C 432&&&&55.5$\pm$ 11.1&42.0  $\pm$ 7.0&&65.6  $\pm$ 8.3&&&&&\\
&&&&L80&L80&&L80&&&&&\\
3C 454.0&&&24.0 $\pm$ 3.0 &31.9  $\pm$ 9.6&&&&&&&&\\
&& &A68 &L80 &&&&&&&&  \\    

\hline
\end{tabular}
}
\\
References:
A68 -     \citet{1968Afz.....4..129A}, 
B69 -     \citet{1969MNRAS.143..289B}, 
B70 -     \citet{1970ApL.....5..129B}, 
C70 -     \citet{1970A&AS....1..281C}, 
C72 -     \citet{1972A&AS....7....1C}, 
C73 -     \citet{1973A&AS...11..291C}, 
C83 -     \citet{1983AJ.....88...20C}, 
C98 -     \citet{1998AJ....115.1693C}, 
C04 -     \citet{2004ApJS..150..417C}, 
D96 -     \citet{1996AJ....111.1945D}, 
F74 -     \citet{1974A&AS...18..147F}, 
F85 -     \citet{1985A&AS...59..255F}, 
G67 -     \citet{1967MmRAS..71...49G}, 
G81 -     \citet{1981AJ.....86.1306G}, 
G91 -     \citet{1991ApJS...75.1011G}, 
H90 -     \citet{1990MNRAS.246..256H}, 
H93 -     \citet{1993MNRAS.263...25H}, 
H95 -     \citet{1995MNRAS.274..447H}, 
K69 -     \citet{1969ApJ...157....1K}, 
K73 -     \citet{1973AJ.....78..828K}, 
K81 -     \citet{1981A&AS...45..367K}, 
L80 -     \citet{1980MNRAS.190..903L}, 
L81 -     \citet{1981MNRAS.194..693L}, 
M05 -     \citet{2005A&A...435..863M}, 
P66 -     \citet{1966ApJS...13...65P}, 
S73 -     \citet{1973AuJPh..26...93S}, 
SJ95 -    \citet{1995A&AS..113..409S}, 
S95 -     \citet{1995AuJPh..48..143S}, 
V75 -     \citet{1975AJ.....80..931V}, 
W75 -     \citet{1975AuJPA..38....1W}, 
W90 -     \citet{1990PKS...C......0W}, 
W96 -     \citet{1996MNRAS.282..779W}, 
C07-      \citet{2007AJ....134.1245C}

\end{minipage}
\end{sidewaystable}

\end{document}